\documentclass[leqno,12pt,a4paper]{article}
\usepackage{latexsym}
\usepackage{amsmath}
\usepackage{amssymb}
\usepackage{bbm}
\usepackage{amsthm}
\usepackage[misc]{ifsym}

\usepackage[]{algorithm,algpseudocode}
\algnewcommand{\IIf}[1]{\State\algorithmicif\ #1\ \algorithmicthen}
\algnewcommand{\IElse}{\unskip\ \algorithmicelse\ }
\algnewcommand{\EndIIf}{\unskip\ \algorithmicend\ \algorithmicif}

\usepackage{graphicx}
\usepackage{epsfig}

\usepackage{tikz}
\usetikzlibrary{shapes,arrows}

\allowdisplaybreaks

\usepackage{tikz}
 
\newcommand{\boldface}[1]{\boldsymbol{#1}}
\newcommand{\bfa}{\boldface{a}}
\newcommand{\bfb}{\boldface{b}}

\newcommand{\bff}{\boldface{f}}

\newcommand{\bfl}{\boldface{l}}

\newcommand{\bfn}{\boldface{n}}

\newcommand{\bfr}{\boldface{r}}

\newcommand{\bft}{\boldface{t}}
\newcommand{\bfu}{\boldface{u}}

\newcommand{\bfx}{\boldface{x}}

\newcommand{\bfA}{\boldface{A}}

\newcommand{\bfC}{\boldface{C}}

\newcommand{\bfF}{\boldface{F}}

\newcommand{\bfI}{\boldface{I}}

\newcommand{\bfS}{\boldface{S}}

\newcommand{\bfX}{\boldface{X}}

\newcommand{\calD}{\mathcal{D}}

\newcommand{\calG}{\mathcal{G}}
\newcommand{\calH}{\mathcal{H}}

\newcommand{\calN}{\mathcal{N}}

\usepackage{bbold}

\newcommand{\be}{\begin{equation}}
\newcommand{\ee}{\end{equation}}
\newcommand{\bea}{\begin{eqnarray}}
\newcommand{\eea}{\end{eqnarray}}
\newcommand{\bes}{\begin{equation*}}
\newcommand{\ees}{\end{equation*}}
\newcommand{\beas}{\begin{eqnarray*}}
\newcommand{\eeas}{\end{eqnarray*}}
\newcommand{\D}{\displaystyle}
\newcommand{\dd}{\ \mathrm{d}}
\newcommand{\damvar}{\alpha}

\newcommand{\intO}{\int_\Omega}
\newcommand{\intPO}{\int_{\partial\Omega}}

\newcommand{\tr}[1]{\mathrm{tr}[#1]}

\newcommand{\der}[1]{#1_{\partial}}
\newcommand{\lap}{\mathop{}\!\mathbin\bigtriangleup}
\newcommand{\ink}[1]{#1_{\Delta}}
\newcommand{\inkx}{\ink{X}}
\newcommand{\inky}{\ink{Y}}
\newcommand{\inkz}{\ink{Z}}

\newcommand{\pf}[2]{\frac{\partial #1}{\partial #2}}
\newcommand{\comp}[1]{{(#1)}}
\newcommand{\eli}{\comp{e}}

\newcommand{\elk}{\comp{k}}

\newcommand{\diss}{\Pi^\text{diss}}

\newcommand{\Nik}{\calN^\eli_k}

\RequirePackage{relsize}
\newfont{\Sf}{cmssbx10 scaled 2074}

\newcommand{\tth}{${}^\text{th}$ }

\usepackage{lipsum}
\usepackage{placeins}

\definecolor{RUBblue}{rgb}{0.0470588,0.262745,0.411765}
\definecolor{blau}{rgb}{0,0.25,1.0}
\definecolor{lightgray}{gray}{0.92}

\newcommand{\coljr}[1]{\bgroup\color{blue}{#1}\egroup}

\usepackage{booktabs}
\newcommand{\titem}{~~\llap{\textbullet}~~}

\begin{document}

\title{Efficient and robust numerical treatment of a gradient-enhanced damage model at large deformations}
\date{} 
\maketitle
{\large
\noindent\underline{Philipp Junker}, Johannes Riesselmann, Daniel Balzani\\[0.5mm]
}
Chair of Continuum Mechanics, Ruhr University Bochum, Bochum, Germany\\[2mm]
{
Corresponding author:\\[0.5mm]
Philipp Junker, \color{RUBblue}{\Letter \hskip 1mm philipp.junker@rub.de} 
}

\section*{Abstract}
The modeling of damage processes in materials constitutes an ill-posed mathematical problem which manifests in mesh-dependent finite element results. The loss of ellipticity of the discrete system of equations is counteracted by regularization schemes of which the gradient enhancement of the strain energy density is often used. In this contribution, we present an extension of the efficient numerical treatment, which has been proposed in \cite{junker2019fast}, to materials that are subjected to large deformations. Along with the model derivation, we present a technique for element erosion in the case of severely damaged materials. Efficiency and robustness of our approach is demonstrated by two numerical examples.

%%%%%%%%%%%%%%%%%%%%%%%%%%%%%%%%%%%%%%%%%%%%%%%%%%%%%%%%%%%%%%
\section{Introduction}
The modeling of damage at finite strains roots back to the eighties of the 20$^\text{th}$ century when first phenomenological models related to material degradation were developed. These models were mainly constructed based on the original ideas for small strains where damage was considered as a reduction of the effective cross section area on which the load acts resulting from the nucleation and evolution of cavities. The reduced cross section thus softens the material which gives rise to the stress- and strain-softening effects. On a material point level, this means that the stress reduces with increasing strains after a critical threshold value is reached. This phenomenological approach of describing the complex microstructural processes during material failure, that might be accompanied by the evolution of dislocations, micro-cracks and delamination effects, provides a convincing method for material modeling. Unfortunately, softening may also be accompanied by a loss of ellipticity of the system of governing equations once a certain amount of damage has evolved which renders boundary value problems ill-posed. An obvious indicator for such problems are severe dependencies of the finite element results on the individual mesh discretization: the smaller the finite element size is chosen, the more localized resulting damage bands will be. Furthermore, and even worse, the numerical solution of such problems does not converge and might be unstable regarding small perturbations of the boundary conditions, i.e., they lack numerical robustness. These effects are not only observed at the level of spatial distribution of the damage variable but also for the global behavior in terms of reaction forces \cite{bavzant2002nonlocal,peerlings2001critical}. Consequently, a mathematical correction of the problem is inevitable. This process is referred to as regularization which might be performed in time and space. 
Overview articles may be found, e.g., in \cite{forest2004localization} and~\cite{yang2012micromechanics}.

Alternatively, a relaxed variational formulation might be employed  
which relies on a careful analysis of the convexity properties of the underlying energy.
This energetic analysis demands a direct formulation of the damage variable as pure function of the deformation state such that the condensed energy (energy plus time-integrated dissipation) may be relaxed. It is thus convenient to formulate the model in a time-incremental way, cf.~\cite{ortiz1999variational,mielke2003energetic,hackl2003dissipation,miehe2002strain}. Thereby, a dissipative material behavior may be described by an incrementally hyperelastic representation, which is thus open for analysis with respect to generalized convexity conditions. This enables the construction of convex hulls (see \cite{gurses2011evolving}  for small strains and \cite{balzani2012relaxed} and \cite{schmidt2016relaxed} for finite strains) which on the one hand ensures the recovery of ellipticity and on the other hand guarantees the existence of minimizers.
In contrast to spacial regularization approaches, no internal length scale parameter has to be chosen since the micromechanical parameters are obtained as result of the convexification procedure. On the other hand, the effort at the material point level may be increased if a nonconvex minimization problem is solved as part of the convexification at each integration point, which yields an increased expense in the assembling procedure.  Another drawback may be seen in the fact that although the stress softening hysteresis as explained above may in principle be captured, a softening in the sense of negative tangents in stress-strain curves can not yet be modeled at the material point. One exception so far is the approach of Schwarz~\textit{et al.} in \cite{schwarz2020variational} for small strains, where a simplified microstructure consisting of damaged, linear elastic materials and whose resulting convexified energy indeed enables the description of strain-softening.

In the context of time-wise regularization, viscous effects are included into the model. Examples for such models may be found in \cite{faria1998strain,Needleman1988,suffis2003damage,forest2004localization,Junker2017relaxation-damage,langenfeld2018quasi}. These models introduce a rate-dependence of the damage evolution which damps the local reduction of stiffness. Consequently, damage evolves also at neighbored material points to keep the amount of dissipated energy constant. A huge benefit of such models (as well as of relaxed incremental formulations) is their implementability in any finite element framework using solely the material law interface. However, viscous damage models possess a remarkable drawback: the viscosity associated to damage evolution has to be chosen quite high, reaching values that are unphysical, in order to obtain the desired regularization. Furthermore, the effectiveness of viscous regularization strongly depends on the applied loading velocity which does not necessarily need to agree with experimental velocities. Finally, the interplay of viscosity and loading velocity has to be set and thus well-posedness has to be proven for each individual boundary value problem rendering such approaches rather impractical for application.

Spacial regularization ``homogenizes'' the damage evolution on a local level. To this end, either integral formulations are introduced or a gradient enhancement of the strain energy is employed. A major benefit of such models as compared to viscous regularization is that it can be shown that they are inherently well-posed. Examples for spatially regularized damage models may be found, e.g., in \cite{bavzant2002nonlocal,peerlings2001critical,borst2001some,PeeBorBreVre:1996:ged,PeeMasGee:2003:atm,dimitrijevic2008method,gross2017fracture,kiefer2018gradient}. 
Also for the finite-strain regime, they have been used, cf. \cite{miehe2016phase,ambati2016phase,carollo20173d,borden2016phase,gultekin2016phase,fathi2017finite,WafPolMenBla:2013:age,brepols2017gradient}. Such models may be applied to arbitrary loading velocities since any dependence of the material behavior on this boundary condition is included in a physically sound manner. Furthermore, they might be coupled to other physical processes as for instance plasticity and temperature dependence without unphysical parameters being present in the complete system of model equations. 
For gradient regularization, we refer to the prototype approach of \cite{PeeBorBreVre:1996:ged,PeeMasGee:2003:atm} and similarly \cite{dimitrijevic2008method}, and its
 finite-strain extension in \cite{WafPolMenBla:2013:age}, where a nonlocal function carries information on the damage state. This procedure transforms the associated equation for the damage state from a local evolution equation at the material point (as is the case for viscous regularization) to a field equation (integral equation for integral regularization or partial differential equation for gradient regularization) which has to be solved along with the balance of linear momentum for the displacements. Consequently, the number of unknowns increases during numerical treatment, e.g., the number of nodal variables increases from three (displacements) to four (displacements + damage state variable) in a mixed finite element implementation. 
While thermodynamic consistency and mesh-independent simulations can be ensured with these approaches, unfortunately, the increased number of degrees of freedom in the discrete global system of equations significantly increases computational costs.

To limit the usual numerical deficiency of gradient regularization, we present in this contribution a novel finite strain gradient damage formulation. 
A similar approach for the small strain regime has been introduced in \cite{junker2019fast} (see also \cite{vogel2020adaptive} for a detailed numerical analysis including mesh adaptivity). 
We begin with the variational derivation of the damage model that is formulated in terms of a partial differential inequality which complements the partial differential equation describing the balance of linear momentum. Then, we apply the neighbored element method which has proven to be beneficial for a robust and time-efficient numerical evaluation of the set of governing equations given by the balance of linear momentum and an additional field equation. Examples have been given in the context of small-strain damage \cite{junker2019fast,vogel2020adaptive} and small-strain \cite{jantos2019accurate} and finite-strain topology optimization \cite{junker2020new}. The neighbored element method combines a finite element approach for the balance of linear momentum and a finite difference approach for the additional field equation along with operator splitting techniques. Thereby, the neighbored element method conserves the size of the global system of equations resulting from the finite element formulation for the displacements: to be more precise, the number of nodal unknowns for the finite elements remains unchanged from a corresponding purely elastic formulation. The partial differential inequality for the damage state, however, is solved through a correspondingly modified internal update procedure. Furthermore, a novel stabilization technique based on element erosion methods enables us to obtain numerical results even for substantial deformation states including regions of highly damaged material for which the model is close to the loss of its validity. The numerical results show mesh-independence while requiring minimal extra computation time compared to purely hyperelastic simulations.

%%%%%%%%%%%%%%%%%%%%%%%%%%%%%%%%%%%%%%%%%%%%%%%%%%%%%%%%

\section{A gradient-enhanced damage model at finite strains}
There exist various strategies for the derivation of fundamental material models of which an extended Hamilton principle offers a variational approach. Its stationarity conditions agree with the 2$^\text{nd}$ Law of Thermodynamics and Onsager's principle by construction if physically reasonable ansatzes for the strain energy density and the dissipation function are chosen. For details on the extended Hamilton principle and its relation to thermodynamics and other modeling strategies, such as the principle of virtual work or the principle of the minimum of the dissipation potential, we refer to \cite{JuBa20-Hamilton}.

The extended Hamilton principle is related to the stationarity of the action function. It can be shown, cf.~\cite{JuBa20-Hamilton}, that it constitutes as the following condition: for the quasi-static case, the sum of the Gibbs energy $\calG$, which is also referred to as total potential, and the work due to dissipative processes $\calD$ tends to be stationary:
\be
\label{eq:Hamilton1}
\calH[\bfu,\alpha] := \calG[\bfu,\alpha] + \calD[\alpha] \rightarrow \underset{\bfu,\alpha}{\text{stat}} \ .
\ee
A detailed investigation of the extended Hamilton principle as unifying theory for coupled problems, i.e. thermo-mechanical processes, and dissipative microstructure evolution has been presented in \cite{JuBa20-Hamilton}.
 
In \eqref{eq:Hamilton1}, the displacements are denoted by $\bfu$ and the state of microstructure is expressed in terms of the internal variable $\alpha$. The Gibbs energy reads
\be
\calG[\bfu,\alpha] = \intO \Psi(\bfC,\alpha) \dd V - \intO \bfb^\star \cdot \bfu \dd V - \intPO \bft^\star\cdot\bfu \dd A
\ee
with the strain energy density $\Psi$, the prescribed body forces $\bfb^\star$, and the traction vector $\bft^\star$. The integrals are evaluated for the body's volume $\Omega$ and its surface $\partial\Omega$ in its reference configuration with the position vector $\bfX$. The deformation is measured by the right Cauchy Green tensor $\bfC := \bfF^\mathrm{T}\bfF$ with the deformation gradient $\bfF = \partial\bfx/\partial\bfX=\bfI + \bfu\otimes\nabla$ and the spatial coordinate $\bfx=\bfX+\bfu$ in the current configuration. Throughout this contribution, the nabla operator is computed with respect to the reference configuration, i.e., $\nabla \equiv \nabla_{\bfX}$.

The work due to dissipative processes is given by
\be
\calD[\alpha] := \intO p^\text{diss} \, \alpha\dd V
\ee
when the non-conservative force $p^\text{diss}$ performs work ``along'' the microstructure state which is described in terms of the internal variable $\alpha$. The stationarity condition of \eqref{eq:Hamilton1} thus reads
\be
\delta\calH[\bfu,\alpha](\delta\bfu,\delta\alpha) = \delta\calG[\bfu,\alpha](\delta\bfu,\delta\alpha) + \delta\calD[\alpha](\delta\alpha) = 0 \qquad \forall \ \delta\bfu, \delta\alpha
\ee
with
\be 
\delta\calG[\bfu,\alpha](\delta\bfu,\delta\alpha) = \intO \Big( \delta_{\bfu} \Psi(\bfC,\alpha) \dd V +  \delta_{\alpha} \Psi(\bfC,\alpha) \Big) \dd V -  \intO \bfb^\star \cdot \delta\bfu \dd V - \intPO \bft^\star\cdot\delta\bfu \dd A \ .
\ee
The non-conservative force is assumed to be a material-specific quantity which is determined once the process conditions are set, similarly as the external forces $\bfb^\star$ and $\bft^\star$. Consequently, it does not follow from a stationarity condition but it needs to be modeled. This implies $\delta_\alpha p^\text{diss}=0$ such that the variation of the dissipated energy reads
\be
\delta\calD[\alpha](\delta\alpha) = \intO p^\text{diss} \,  \delta\alpha\dd V \ .
\ee
Since the variations for displacements and microstructural state are independent, we obtain
\be
\label{eq:stationarity1}
\begin{cases}
\D \intO  \delta_{\bfu} \Psi(\bfC,\alpha) \dd V -  \intO \bfb^\star \cdot \delta\bfu \dd V - \intPO \bft^\star\cdot\delta\bfu \dd A &= 0 \qquad \forall \ \delta\bfu \\[5mm]
\D \intO \delta_{\alpha} \Psi(\bfC,\alpha) \dd V + \intO p^\text{diss} \, \delta\alpha \dd V &=  0 \qquad \ \forall\delta\alpha
 \end{cases}  \ .
\ee
Specifications of the strain energy density $\Psi$ and the non-conservative force $p^\text{diss}$ allow the application of the latter equations for the derivation of models for various materials. Based thereon, let us provide specific formulas for a gradient-enhanced damage model by first setting the internal variable $\alpha$ as damage variable. Then, we define the strain energy density by
\be
\Psi(\bfC,\damvar) = \big(1-D(\damvar)\big)\, \Psi_0(\bfC) + \frac{1}{2} \beta ||\nabla f||^2
\ee
with the damage function $D(\damvar):=1-f(\damvar)$, the damage variable $\alpha$ and  $f(\damvar)=\exp(-\damvar)$. The parameter $\beta$ serves as regularization parameter and thus controls the thickness of the damage zone. The formulation of the strain energy density is the same approach as in \cite{junker2019fast}; however, in contrast to the original work which was restricted to small deformations, we now make use of a hyperelastic strain energy density of the undamaged material $\Psi_0=\Psi_0(\bfC)$. 

The non-conservative force is usually derived from a dissipation function $\diss$ such that
\be
\label{eq:pDissPartial}
p^\text{diss} = \pf{\diss}{\dot{\alpha}}
\ee
holds. For rate-independent microstructural evolution, which is the case for the present model for damage evolution, functions have to be used that are homogeneous of order one. Hence, the dissipation function
\be
\label{eq:DefDiss}
\diss := r |\dot{\damvar}|
\ee
is used \cite{junker2019fast} from which 
\be
p^\text{diss} = \partial\diss := \begin{cases} \D  r \frac{\dot{\damvar}}{|\dot{\damvar}|} & \text{for } \dot{\alpha}\not= 0 \\ \D  \{-r,r\} & \text{for } \dot{\alpha}= 0 \end{cases}
\ee
follows. The parameter $r$ is referred to as dissipation parameter and it will be become obvious that it represents an energetic threshold value for the onset and evolution of damage. It is worth mentioning that the partial derivative in~\eqref{eq:pDissPartial} transforms to a subdifferential in our case since $\partial\diss/\partial\dot{\alpha}$ is not unique at $\dot{\alpha}=0$ for our rate-independent ansatz for $\diss$ in \eqref{eq:DefDiss}. Then, the stationarity condition in \eqref{eq:stationarity1}$_1$ result in the weak form of the balance of linear momentum
\be
\intO \bfS: \frac{1}{2}\delta\bfC \dd V  -  \intO \bfb^\star \cdot \delta\bfu \dd V - \intPO \bft^\star\cdot\delta\bfu \dd A = 0 \qquad \forall \ \delta\bfu \ ,
\ee
where $\bfS=2\partial\Psi(\bfC,\damvar)/\partial\bfC$ denotes the 2$^{\text{nd}}$ Piola Kirchhoff stress tensor. The stationarity condition in \eqref{eq:stationarity1}$_2$ reads
\be
\label{eq:StatF}
\intO f^\prime \Psi_0 \, \delta\alpha \dd V + \intO \beta \nabla f \cdot \nabla( f^\prime \delta \alpha) \dd V + \intO \partial\diss \delta\alpha \dd V = 0 
\ee
where we considered $\delta_\alpha f = f^\prime \delta\alpha$. Integration by parts of the second term results in
\be
\intO \beta f^\prime \nabla f \cdot \nabla \delta \alpha \dd V = \intPO \beta f^\prime \bfn_0 \cdot \nabla f \, \delta\alpha \dd A - \intO \beta f^\prime   \lap f \, \delta\alpha \dd V
\ee
and thus, \eqref{eq:StatF} transforms to
\be
\label{eq:StatF2}
- \intO \left(f \Psi_0 - \beta f \lap f - \partial\diss \right) \delta\alpha \dd V - \intPO \beta f \bfn_0 \cdot \nabla f \, \delta\alpha \dd A = 0
\ee
due to $f^\prime=-f$. The local evaluation of the first integral in \eqref{eq:StatF2} results in the differential inclusion
\be
f \Psi_0 - \beta f \lap f - \partial\diss \ni 0 \quad \forall \bfX \in \Omega
\ee
due to the set-valued character of the subdifferential $\partial\diss$. Therefore, it is convenient to perform a Legendre transformation such that we transform the dissipation function, which is formulated in terms of the thermodynamic flux $\dot{\alpha}$, into a function $\tilde\Pi^{\mathrm{diss}}$ which depends on the thermodynamic force $p:=f \Psi_0 - \beta f \lap f$. Consequently, we obtain
\be
\tilde\Pi^{\mathrm{diss}} = \underset{\dot{\alpha}}{\mathrm{sup}} \Big\{  p \dot{\alpha} - \diss \Big\} = \underset{\dot{\alpha}}{\mathrm{sup}} \Big\{ |\dot{\alpha}|(p \,\mathrm{sgn}\dot{\alpha} - r) \Big\} \ .
\ee
Healing in the sense of an increasing stiffness contradicts our motivation to model damage. Thus, the sign of the rate of the damage variable is $\mathrm{sgn}\dot{\alpha}=\{0,1\}$ and the Legendre transform reads
\be
\tilde\Pi^{\mathrm{diss}} = \begin{cases} 0 & \text{if } p - r \le 0  \\ \infty  & \text{else} \end{cases}
\ee
From the first case, we identify
\be
\begin{cases}
p<r : & \dot{\alpha} = 0 \\
p=r : & \dot{\alpha} > 0
\end{cases} \ .
\ee
It is thus convenient to introduce the indicator function $\Phi:= p -r $ which separates stationarity of the damage state for $\Phi<0$ and evolution of the damage state for $\Phi=0$. The indicator function fulfills a similar purpose as yield functions in models for elasto-plasticity. The parameter $r$ has a similar meaning as a scalar-valued yield stress. However, it is an energetic threshold value for the present case of damage. Finally, we can collect the governing equations for damage evolution in the following form
\be
\label{eq:KKT}
\dot{\damvar} \ge 0 \ , \qquad \Phi := f \Psi_0 - \beta f \lap f - r \le 0 \ , \qquad \Phi \,\dot{\damvar} = 0 \qquad \forall \ \bfX \in\Omega 
\ee
which are identified as Karush Kuhn Tucker conditions. The surface integral constitutes as Neumann condition 
\be
\label{eq:Neumann}
\bfn_0\cdot\nabla f = 0 \qquad \forall \ \bfX \in \partial\Omega
\ee
for $f$. Here, $\bfn_0$ denotes the normal vector on the surface $\partial\Omega$ in the reference configuration. More details on the fundamentals of the damage model can be found in the original publication \cite{junker2019fast}.

\section{Numerical treatment}
\label{sec:Numerical}
The model above consists of two unknowns: the displacement field $\bfu$ and the damage variable $\alpha$. They can be determined by solving 
\be
\label{eq:System}
\begin{cases}
\D \intO \bfS: \frac{1}{2}\delta\bfC \dd V  -  \intO \bfb^\star \cdot \delta\bfu \dd V &\D  - \intPO \bft^\star\cdot\delta\bfu \dd A = 0 \qquad \forall \ \delta\bfu \\[4mm]
& \D \  f \Psi_0 - \beta f \lap f - r \le 0 \qquad \forall \ \bfx\in\Omega
\end{cases}
\ee
with the Dirichlet and Neumann boundary conditions $\bfu=\bfu^\star \ \forall \bfX\in\partial\Omega_u$ and $\bfF\bfS\bfn_0=\bft^\star \ \forall\bfX\in\partial\Omega_\sigma$ with $\partial\Omega=\partial\Omega_u\cup\partial\Omega_\sigma$ and $\partial\Omega\cap\partial\Omega_\sigma=\emptyset$ for \eqref{eq:System}${}_1$. For \eqref{eq:System}${}_2$, the initial condition $f(t=0)=1 \ \forall\bfX\in\Omega$ and the Neumann condition $\bfn_0\cdot\nabla f=0 \ \forall\bfX\in\partial\Omega$ apply.

From \eqref{eq:System}${}_2$, we recognize that a partial differential inequality has to be solved for the primal variable $f$. Once $f$ is determined, the value of the damage variable could be computed subsequently for each material point $\bfX$ by $\damvar=-\log[f(\bfX)]$; however, this information is of minor interest. More importantly, the damage function $D=1-f$ can be computed which governs the stresses by
\be
\label{eq:2ndPKdamage}
\bfS = (1-D) \bfS_0 \qquad \text{with} \qquad \bfS_0 := \pf{\Psi_0}{\bfC} \ .
\ee
The partial differential inequality in \eqref{eq:System}$_2$ is rather untypical in the context here and a standard finite element treatment would be complex due to the subdifferential in its weak form, cf.~\eqref{eq:StatF}. Furthermore, additional degrees of freedom at the nodes would be present in the numerical solution scheme. To avoid both drawbacks, we follow \cite{junker2019fast} and make use of the neighbored element method: standard finite element approaches are applied to the weak form of the balance of linear momentum whereas a finite difference approach for unstructured grids is used for discretizing the strong form of the evolution equation for $f$. The neighbored element method is completed by utilization of an operator split, i.e., the discretized equations are solved in a staggered manner. This procedure of combining staggered FEM and FDM has also been proven advantageous for the small-strain regime in \cite{junker2019fast}. In \cite{vogel2020adaptive}, it has been demonstrated that convergence is achieved which justifies the operator split. The finite element treatment of the nonlinear weak form of the balance of linear momentum is standard such that we dispense with a detailed presentation. More details on the finite element method can be found in standard textbooks as in, e.g. \cite{wriggers2008nonlinear}. The only modification is, of course, that both the stress and the stiffness are scaled by the damage state, cf. \eqref{eq:2ndPKdamage}.

For the derivation of an appropriate finite difference method that also operates on unstructured grids, we employ a Taylor series expansion up to order two in order to approximate $f$. Usually, the value of the function, i.e., its zeroth derivatives, along with the derivatives of higher order are known when a Taylor series expansion is used. Then, due to the chosen spatial increment, the value of the function can be approximated at a neighbored spatial point. This operation is inverted here: the value of the damage function at specific points are all known. These points are the centers of gravity of all finite elements. Then, the spatial increments in the Taylor series expansion are given by the spatial increments of the centers of gravity; they remain fixed if no mesh adaptation is employed during the computation. Consequently, the only unknown in the Taylor series expansion are the (mixed) derivatives of order $\ge 1$. Collecting an appropriate set of neighbored elements, a linear system of algebraic equations can be constructed to compute the individual (partial) derivatives from which the local value of the Laplace operator follows by simple summation of the unmixed derivatives of order two. 

Our finite differences approach at unstructured meshes can be recast in mathematical formulas by first introducing the Taylor series expansion as
\be
\label{eq:Taylor1}
f^{(\calN^\eli_k)} = f^\eli  + \sum_o  A^{\comp{i,k}}_o f^\eli_{\partial,o}
\ee
where $f^\eli_{\partial,o}$ stores the value of the partial derivatives of varying order $o\ge1$, such that a matrix including all derivatives can be defined as 
\be
\label{eq:fpartial}
\der{\bff}^\eli := \left( \begin{matrix}
\D \frac{\partial f^\eli}{\partial X} & \D \frac{\partial f^\eli}{\partial Y} &  \D \frac{\partial f^\eli}{\partial Z} & \D\frac{\partial^2 f^\eli}{\partial X \partial Y} & \D\frac{\partial^2 f^\eli}{\partial Y \partial Z} & \D\frac{\partial^2 f^\eli}{\partial X \partial Z} &  \D\frac{\partial^2 f^\eli}{\partial X^2} & \D\frac{\partial^2 f^\eli}{\partial Y^2} & \D\frac{\partial^2 f^\eli}{\partial Z^2}
 \end{matrix} \right) \ .
\ee
The quantity $A^{\comp{i,k}}_o$ comprises all spatial increments with varying power such that again the associated column matrix is defined as 
\bea
 A^\comp{k} & :=  &  \Bigg(
\inkx^\comp{\Nik} \quad \inky^\comp{\Nik} \quad \inkz^\comp{\Nik} \quad 
\inkx^\comp{\Nik} \inky^\comp{\Nik} \quad \inky^\comp{\Nik} \inkz^\comp{\Nik} \quad \inkx^\comp{\Nik}\inkz^\comp{\Nik}  \\
%\ink{(XY)}^\comp{\Nik} \quad \ink{(YZ)}^\comp{\Nik} \quad \ink{(ZX)}^\comp{\Nik}  \\
& &  \qquad\qquad\qquad\qquad\quad   \frac{1}{2}\Big( \inkx^\comp{\Nik}\Big)^2 \quad \frac{1}{2}\Big( \inky^\comp{\Nik}\Big)^2 \quad \frac{1}{2}\Big( \inkz^\comp{\Nik}\Big)^2 
\Bigg) \notag
\eea
with
\begin{align*}
\inkx^\comp{\Nik} := X^\comp{\Nik}-X^\eli \ , \quad %\ , \qquad  & \ink{(XY)}^\comp{\Nik} := (X^\comp{\Nik}-X^\eli)(Y^\comp{\Nik}-Y^\eli) \\[1mm]
\inky^\comp{\Nik} := Y^\comp{\Nik}-Y^\eli \ , \quad %\ , \qquad & \ink{(YZ)}^\comp{\Nik} := (Y^\comp{\Nik}-Y^\eli)(Z^\comp{\Nik}-Z^\eli) \\[1mm]
\inkz^\comp{\Nik} := Z^\comp{\Nik}-Z^\eli \ . %\ , \qquad &  \ink{(ZX)}^\comp{\Nik} := (Z^\comp{\Nik}-Z^\eli)(X^\comp{\Nik}-X^\eli) \ .
\end{align*}
To ensure that only the required number of elements and thus extrapolation points is used, the quantity $\Nik$ has been introduced: $\calN^\eli$ is the set of neighbored elements around element $\eli$ and, hence, $\Nik$ returns the global element number for $k$\tth neighbor. All column matrices $A^\comp{k}$ in the neighborhood $\Nik$ of the element of interest $\eli$ are collected in the matrix $\bfA^\eli$. Then, by defining
\be
f^\eli_{\Delta,k} := f^{(\calN^\eli_k)}-f^\eli \ ,
\ee
we can shortly write for the Taylor series expansion \eqref{eq:Taylor1}
\be
\label{eq:TaylorSystem}
\ink{\bff}^\eli = \bfA^\eli \der{\bff}^\eli 
\ee
such that the unknown vector of derivatives follows from
\be
\der{\bff}^\eli  = \bfA^{\eli-1} \ink{\bff}^\eli \ .
\ee
The Laplace operator can be computed by aid of $\der{\bff}^\eli$ according to
\be
\lap f^\eli = \frac{\partial^2 f^\eli}{\partial X^2} + \frac{\partial^2 f^\eli}{\partial Y^2} + \frac{\partial^2 f^\eli}{\partial Z^2} = f^\eli_{\partial, 7} +  f^\eli_{\partial, 8} +  f^\eli_{\partial, 9} \ ,
\ee
cf. \eqref{eq:fpartial}. Consequently, only some components of $\der{\bff}^\eli$ are of interest. Introducing
\be
\bfa := \left(\begin{matrix} 0 & 0 & 0 & 0 & 0 & 0 & 1 & 1 & 1 \end{matrix} \right) \ ,
\ee
the Laplace operator is simply given by
\be
\label{eq:lap}
\lap f^\eli = \bfl^\eli \cdot \ink{\bff}^\eli
\ee
where the vector
\be
\label{eq:lapM}
\bfl^\eli := \bfa^\mathrm{T}  \bfA^{\eli\,-1} 
\ee
can be computed once for each element $\eli$ in advance of the actual solution of the boundary value problem. In contrast, the evaluation of \eqref{eq:lap} has to be repeated during computational evaluation of \eqref{eq:System}$_2$ to find the (current) field $f$; this evaluation, however, is computationally very cheap. It is worth mentioning that the dimension of $\bfA^\eli$ depends on cardinality of the set of neighbored elements $\calN^\eli$: to close the system of equations in \eqref{eq:TaylorSystem}, i.e., to ensure that $\bfA^\eli$ is a regular matrix, the cardinality of the set of neighbored elements has at least to equal the length of $\der{\bff}^\eli=9$ in the considered three-dimensional case. For boundary elements, however, we usually find a smaller cardinality of $\calN^\eli$, i.e. less than nine neighbored elements can be identified. To circumvent this problem, we introduce ghost elements at the boundary which mirror the value of the damage field inside of $\Omega$ to the outer vicinity of the boundary $\partial\Omega$ which equals the usual treatment in the finite difference method. It is worth mentioning that \eqref{eq:Neumann} is fulfilled identically by usage of ghost elements. For the elements inside of $\Omega$, the opposite case is present: considering all neighbored elements around $\eli$, i.e., the stencil of six elements plus eight diagonal elements plus twelve elements in each ``plane'', results in an overdetermined system of equations for \eqref{eq:TaylorSystem}. Here, two possibilities can be employed: the first option is to take into account element $\eli$ and all 26 neighbored elements and compute the inverse of $\bfA^\eli$ by use of the right-inverse as
\be
\bfA^{\eli\,-1} = \bfA^{\eli\,\text{T}}\left(\bfA^\eli\bfA^{\eli\,\text{T}}\right)^{-1} \ ,
\ee
cf.~\cite{vogel2020adaptive}. The second option is to use the six stencil elements plus three diagonal elements with closest distance to the center element $\eli$ as neighborhood. For sufficiently regular meshes, $\bfA^\eli$ contains full rank and is regular, cf.~\cite{vogel2020adaptive}. In all computational results we present later, we made use of the latter option. We refer to \cite{junker2019fast,jantos2019accurate,vogel2020adaptive} for more details on the neighbored element method.

Finally, it is mandatory to chose a strategy for solving the discretized form of \eqref{eq:System}$_2$ given by
\be
\label{eq:IndicatorElement}
\Phi^\eli=f^\eli \bar{\Psi}_0^\eli - \beta f^\eli \ \bfl^\eli \cdot \ink{\bff}^\eli - r \le 0 
\ee
where we introduced the strain energy density averaged for each finite element as
\be
\bar{\Psi}^\eli_0 := \frac{1}{\Omega^\eli} \int_{\Omega^\eli} \Psi_0 \dd V \ .
\ee
The Laplace operator turns \eqref{eq:IndicatorElement} into a system of coupled inequalities which could be solved in a monolithic manner. Although being the obvious strategy, the coupling demands the usage of (external) solver routines. In combination with the inequalities, this renders such a monolithic strategy inappropriate to be implemented into an arbitrary finite element program. Therefore, a simple yet successful numerical solution procedure has been proposed in \cite{junker2019fast}: the system of coupled inequalities is solved by means of a Jacobi method. To be more precise, the coupling is ignored at first glance and each inequality is checked individually such that an update of $f$ is performed for all elements with $\Phi^\eli>0$. Of course, a modification of $f^\eli$ may have an impact on the indicator function in a different element $\elk$ due to the Laplace operator which ``transports'' the information. Therefore, the update has to be repeated until $\Phi^\eli\le 0 \; \forall \eli$. Usually, such an update scheme is known to be rather time consuming since the numbers of repetitions needed for convergence depends on the number of unknowns which equals the number of finite elements in our case. This should turn this strategy to be more expensive than monolithic solution schemes. However, the damage evolution is limited to only a small number of elements as compared to their total number and thus, an update is mandatory only for a negligible fraction of inequalities. As will be shown by our numerical experiments, the staggered solution scheme is still very fast and simulations involving damage evolution consume hardly more computation time. A similar behavior has already been demonstrated for the small deformation setting of damage modeling in \cite{junker2019fast} and even for an evolutionary method for topology optimization both at small strains in \cite{jantos2019accurate} and at finite strains in \cite{junker2020new}. After the damage function has been updated, the next time step is investigated without checking the impact of the updated damage field on the displacements for the current time step. This might be interpreted as (additional) operator step and may seem questionable whether a reliable result is obtained by this method. However, time steps with vanishing increment are mandatory for an appropriate resolution e.g. of the force/displacement diagram. For this case, the method of operator splits converges. It has been numerically shown in \cite{junker2019fast} that the artificially included pseudo-viscous behavior due to the operator step does not influence the final result on a relevant scale.  Furthermore, it is worth mentioning that a detailed numerical study on the numerical behavior of the damage model for the small deformation setting in \cite{vogel2020adaptive} revealed that almost identical results are obtained when both fields (displacements and damage) are repeatedly updated until a common convergence is achieved. The update of each inequality is performed by a one-step Newton procedure as
\be
f^\eli \leftarrow f^\eli - \frac{\Phi^\eli}{\mathrm{d}\Phi^\eli}
\ee
with
\be
\label{eq:df}
\mathrm{d}\Phi^\eli = \Psi_0^\eli - \beta \bfl^\eli\cdot\ink{\bff}^\eli + \beta f^\eli \ \bfl^\eli\cdot\boldsymbol{1}
\ee
where the length of the vector $\boldsymbol{1}$ equals the cardinality of the set $\calN^\eli$ (9 for our chosen method for defining the neighborhood) and all elements in $\boldsymbol{1}$ are set to one. The one-step Newton method has been proven beneficial due to its smooth convergence behavior for neighbored elements caused by the undershooting. Additionally, this ensures a monotonic decrease of the damage function and thus agrees to \eqref{eq:KKT}$_1$: the indicator function $\Phi^\eli$ is a convex quadratic function in $f^\eli$, and since the update for $f^\eli$ is only performed for $\Phi^\eli>0$, it holds for the derivative $\mathrm{d}\Phi^\eli>0$. The overshooting ensures that the solution converges from above, i.e. $0\leftarrow\Phi^\eli$, and thus $f^\eli$ is monotonically decreasing which implies $\dot{\alpha}\ge0$.

\begin{algorithm}[htb]
\caption{Element erosion strategy}
\label{alg:stabilization}
\begin{algorithmic}[]
\State \For{$1,\dots,n_e$} \Comment{apply to all finite elements}
\State \If{$D^\eli = D_\text{crit}$} \Comment{check damage criterion}\vspace{3mm}
%\State \If{$D^\eli = 1- f^\eli\ge D_\text{crit}$} \Comment{check damage criterion}\vspace{3mm}
%\State set $D^\eli = D_\text{crit}$ and $f^\eli=1-D_\text{crit}$ \Comment{critical bound for the damage field} \vspace{3mm}
\State set $\bfr^\eli = \boldsymbol{0}$ \Comment{eliminate element residual}\vspace{3mm}
\State set $\D\frac{\mathrm{d}\bfr^\eli}{\mathrm{d}\hat{\bfu}^\eli} = \Bigg( \frac{\mathrm{d}r^\eli_i}{\mathrm{d} \hat{u}^\eli_j} \Bigg)= s_\text{crit} (\delta_{ij}) $ \Comment{erode element $\eli$}
\EndIf\vspace{3mm}
\EndFor
\end{algorithmic}
\end{algorithm}

%\subsection{Stabilization technique}
%\label{ssec:stabilization}
The above mentioned numerical procedure is analogous to the one used for the small deformation setting in \cite{junker2019fast}. However, a remarkable difference is accompanied by the large deformation setting: here, severe deformations that occur during damage evolution are correctly described. This, of course, is the primal intention of using the large deformation setting. Unfortunately, this sensitivity might result in numerical instabilities causing $\det \bar{\bfF}^\eli<0$ where $\bar{\bfF}^\eli:=\int_{\Omega^\eli} \bfF \ \mathrm{d}V / \Omega^\eli$ is the averaged deformation gradient in element $\eli$. To be more precise, deformation states are computed that do not correspond to a physically accurate material behavior: once the deformation state has reached some critical threshold, cracks will be present in a real material such that a consideration of the specimen as one continuous body cannot be justified and a description in a continuum mechanical sense is not justified anymore. To circumvent this numerical artifact and also to improve physical accuracy, a stabilization technique in terms of element erosion is applied. To this end, we define a critical value $D_\text{crit}$ when damage modeling is not suitable anymore but rather cracking needs to be considered. To correctly account for this physical behavior in our finite element simulations, we thus modify the element stiffness by setting it to a diagonal matrix with a virtual remaining stiffness value of $s_\text{crit}$ representing an eroded element through which rank deficiency of the affected elements is avoided. Finally, we set the residual forces of this element to zero and no further evolution of the damage value is considered anymore at this point. 

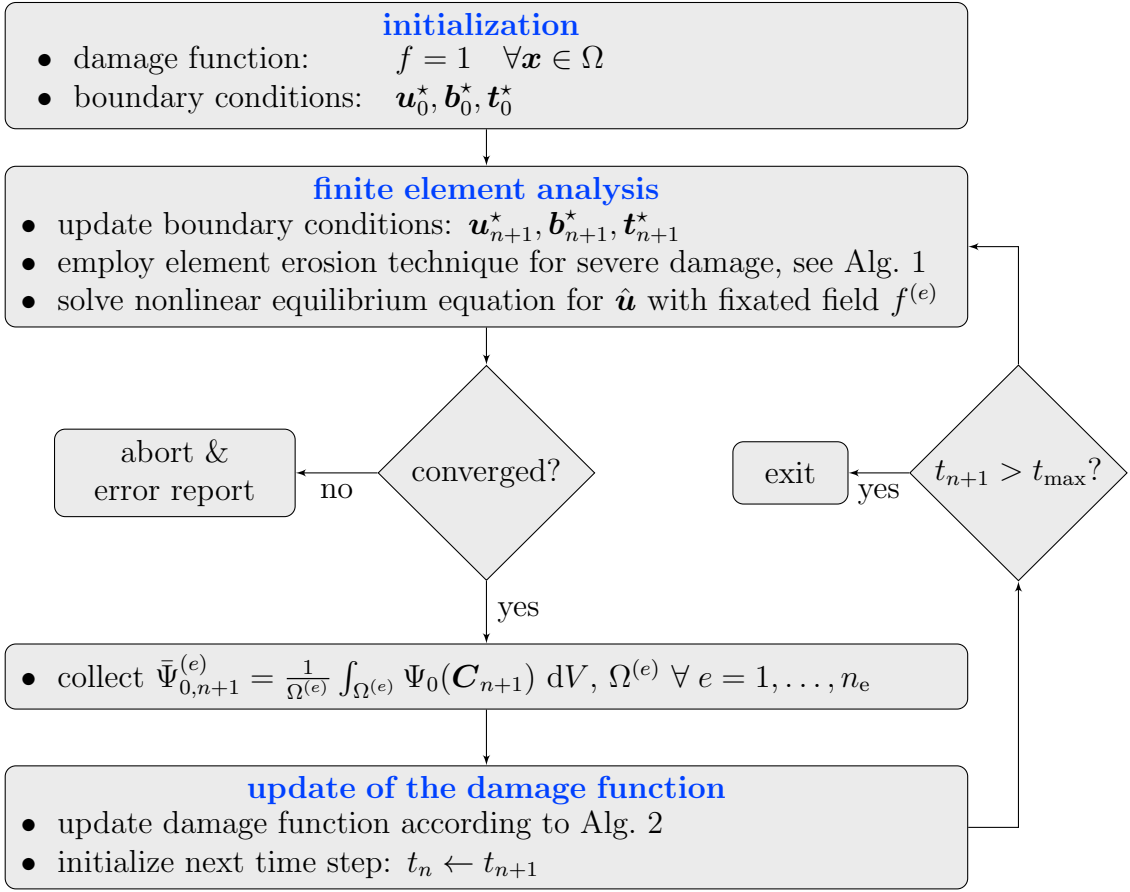
\begin{figure}[ht]
\centering
% Define block styles
\tikzstyle{decision} = [diamond, draw, 
    text width=6em, text badly centered, inner sep=0pt,fill=lightgray]
\tikzstyle{abort} = [rectangle, draw, 
    text width=7.0em, text centered, rounded corners, minimum height=2em,fill=lightgray]    
\tikzstyle{exit} = [rectangle, draw, 
    text width=3.0em, text centered, rounded corners, minimum height=2em,fill=lightgray]    
    \tikzstyle{blind} = [circle, 
    text width=0em, text centered, minimum height=0em]
\tikzstyle{line} = [draw, -latex']
\tikzstyle{ini2} = [rectangle, draw, 
    text width=30em, text centered, rounded corners, minimum height=2em,fill=lightgray]
\tikzstyle{ini22} = [rectangle, draw, 
    text width=30em,  rounded corners, minimum height=2em,fill=lightgray]    
    \begin{tikzpicture}[node distance = 2.5cm, auto]
    % Place nodes
    \node [ini2] (init2) {\textbf{\textcolor{blau}{initialization}} \vspace{-3mm}
\begin{flushleft}
     \begin{tabular}{ll}
      \titem damage function: & $f = 1 \quad \forall \bfx \in \Omega$ \\
      \titem boundary conditions: & $\bfu_0^\star, \bfb_0^\star, \bft_0^\star$
     \end{tabular}
\end{flushleft}     
	};
    \node[ini2, below of=init2, node distance=2.40cm](FE2) { \textbf{\textcolor{blau}{ finite element analysis}}\vspace{-4mm}
\begin{flushleft}  
 \titem update boundary conditions: $\bfu_{n+1}^\star, \bfb_{n+1}^\star, \bft_{n+1}^\star$\\
 \titem employ element erosion technique for severe damage, see Alg.~\ref{alg:stabilization}\\
 \titem  solve nonlinear equilibrium equation for $\hat\bfu$ with fixated field $f^\eli$
\end{flushleft} 
    };
    \node [decision, below of=FE2, node distance=3.0cm] (converged) {converged?};
    \node [abort, left of=converged, node distance=4.1cm] (stop) {abort \& \\ error report};
    \node [blind, below of=converged, node distance=2.0cm] (blind) {};
   \node  [ini22, below of=converged, node distance=2.7cm](update1){  \titem  collect $\bar{\Psi}_{0,n+1}^\eli =\frac{1}{\Omega^\eli} \int_{\Omega^\eli} \Psi_0(\bfC_{n+1}) \ \mathrm{d}V$, $ \Omega^\eli  \ \forall\; e=1,\dots,n_\text{e}$ };
    \node [ini2, below of=update1, node distance=2.0cm] (update2) { \textbf{\textcolor{blau}{update of the damage function}}\vspace{-4mm}
\begin{flushleft}
 \titem update damage function according to Alg.~\ref{alg:damage-update}\\
 \titem initialize next time step: $t_n\leftarrow t_{n+1}$
\end{flushleft}
   };
    \node [blind, right of=FE2,node distance=7cm] (blind3) {};
    \node [decision, right of=converged,node distance=7cm] (converged2) {$t_{n+1}>t_\text{max}?$};
    \node [exit, left of=converged2, node distance=3cm](exit){exit};
        % Draw edges
    \path [line] (init2) -- (FE2);
    \path [line] (FE2) -- node {}(converged);
    \path [line] (converged) -- node {no} (stop);
    \path [line] (converged) -- node {yes} (update1);
    \path [line] (update1) -- node {}(update2);    
    \path [line] (update2.east) -| (converged2);
    \path [line] (converged2) |- (FE2.east);
    \path [line] (converged2) -- node{yes}(exit);
    \end{tikzpicture}
\caption{Flowchart for the proposed damage model.}
\label{fig:Flow}
\end{figure}

Summarizing, we follow the element erosion technique in Alg.~\ref{alg:stabilization} and employ Alg.~\ref{alg:damage-update} for the damage update. The flowchart in Fig.~\ref{fig:Flow} visualizes the neighbored element approach including an interplay between a finite element and finite difference update for the displacements and $f$, respectively.

\begin{algorithm}[htb]
\caption{Damage update}
\label{alg:damage-update}
\begin{algorithmic}[]
\State input $\bar{\Psi}_{0,n+1}^\eli = \int_{\Omega^\eli} \Psi_{0,n+1} \ \mathrm{d}V / \Omega^\eli$ \Comment{averaged energy for each finite element $\eli$}\vspace{2mm}
\State \hphantom{input} $\bar{\bfF}^\eli = \int_{\Omega^\eli} \bfF \ \mathrm{d}V / \Omega^\eli$ \Comment{averaged deformation gradient for each element $\eli$}
\State \For{$1,\dots,n_\text{loop}$} \Comment{Jacobi method: repeat $n_\text{loop}$ times}
\State \State compute $\lap f^\eli$ according to \eqref{eq:lap} \Comment{Laplace operator}
\State \State compute $\Phi^\eli$ according to \eqref{eq:IndicatorElement} \Comment{indicator function}
\State \If{$\Phi^\eli>0$ \textbf{and} $\det \bar{\bfF}^\eli>1$}\vspace{2mm} %\Comment{check damage criterion}
\State update $f^\eli \leftarrow f^\eli - \frac{f^\eli}{\mathrm{d}\Phi}$ according to \eqref{eq:df} \Comment{new damage value for inelastic $\Phi^\eli$}
\State \If{$D^\eli=1-f^\eli > D_\text{crit}$} 
\State set $D^\eli = D_\text{crit}$ and $f^\eli=1-D_\text{crit}$ \Comment{activation of element erosion}
\EndIf \vspace{3mm}
\Else
\State set $\Phi^\eli=0$ \Comment{elimination of elastic $\Phi^\eli$}
\EndIf
\State \If{$\max\{\Phi^\eli\}< 10^{-6}$} 
\State exit \Comment{terminate Jacobi method}
\EndIf
\EndFor
\end{algorithmic}
\end{algorithm}

\section{Numerical Tests \label{s:numerical_tests}}
In this section, the introduced approach is numerically tested in a 3D finite strain setting. To this end, corresponding 8-node hexahedral finite elements were implemented and evaluated within the finite element program FEAP, cf.~\cite{FEAP}. For the strain energy corresponding to the fictively undamaged state, the Neo-Hooke energy given as
%
%neo hooke
\begin{align}
  &\Psi_0 =
   \frac{\mu}{2}(I-3) + g(J)
   \label{e:neo_hooke}
\end{align}
with $g(J) = \frac{\lambda}{4}\left(J^2-1\right)-\left(\frac{\lambda}{4}+\mu\right) \ln J$ and $I=\tr{\bfC}$, $J=\det{\bfF}$ is employed (see also \cite{wriggers2008nonlinear}).
The Lam\'e constants  $\lambda=E\nu/((1+\nu)(1-2\nu))$ and $\mu=E/(2(1+\nu))$ are  computed from the Young's modulus $E$ and the Poisson ratio $\nu$.
Throughout the numerical tests, an element erosion approach is incorporated as discussed in Sec.~\ref{sec:Numerical} through which, once the damage value of an element succeeds a critical value $D^\eli>D_{\text{crit}}$, the element stiffness is replaced by a virtual remaining stiffness $s_\text{crit}$ and residual forces are eliminated, cf. Alg.~\ref{alg:stabilization}.
Moreover, to avoid damage evolution under compression, we add the condition $\det \bar\bfF^\eli>1$
to the damage update algorithm in Alg.~\ref{alg:damage-update}.
For the solution procedure uniform incremental loading is applied and the solution of the linearized system corresponding to the Newton-Raphson iterations is performed with the PARDISO solver. The parameters for the technique for element erosion are collected in~Tab.~\ref{t:parameters}.

% parameters used throughout the tests
\begin{table}
\caption{Material and boundary parameters used throughout the numerical tests.}
\begin{center}
\begin{tabular}{lllll} \toprule
%  \multicolumn{3}{c}{\textbf{Material Data}} \\
Description                    &    Symbol                  & \multicolumn{2}{c}{Value}                  & Unit     \\ \cmidrule(lr){3-4}
                               &                      & Plate with hole                &   U-shape       &     \\ \midrule    %  & Description                    
E-modulus                      &$E $                  &    $500$                       &   $1000$        &      MPa           \\  %             & Elasticity modulus  
poisson ratio                  &$\nu$                 &   $0.3$                        &   $0.3$         &     -   \\  %                 & Poisson ratio   
dissipation parameter          &$r$                   &   $5.0$                        &   $0.5$         & MPa                  \\  %             &   Dissipation parameter 
nonlocal parameter             &$\beta$               &    $\in\{10,100,1000\}$        &   $100$         & N    \\  %      &     Nonlocal parameter       
critical damage value           &$D_{\mathrm{crit}}$            &    $0.95$                      &   $0.995$       & -                 \\    
critical stiffness & $s_\text{crit}$   & $10^{-8}$ & $10^{-8}$ & N/mm \\ 
load increment                 & -         &   $\{0.25,0.1,0.025,0.01\}$ &   $0.2$         & mm    \\ \bottomrule
\end{tabular} 
\end{center}
\label{t:parameters}
\end{table}

\subsection{Plate with hole benchmark test \label{ss:plate_with_hole}}
%% 3d plate with hole description 
\begin{figure}
\unitlength1cm
% \begin{center}
\begin{minipage}{0.38\textwidth}
(a)
 \def\svgwidth{\textwidth}
  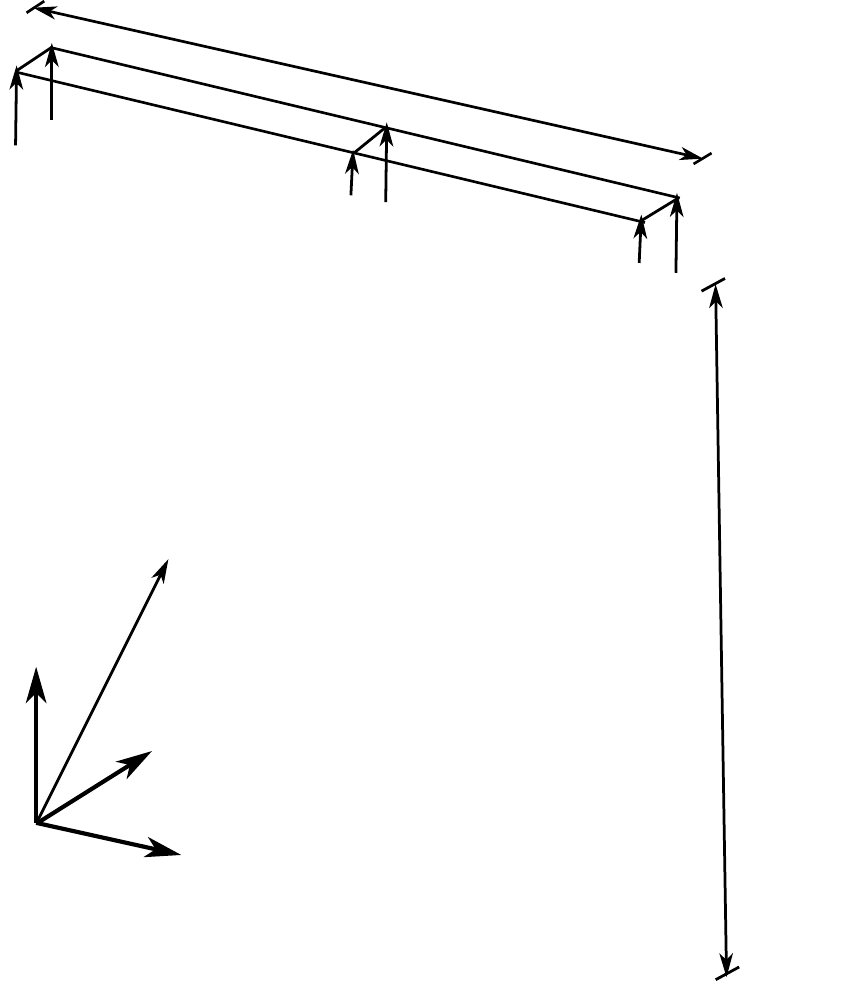
%   (a)
\end{minipage}
\begin{minipage}{0.6\textwidth}
\begin{picture}(0,7)
\put(0,0.5){
  \includegraphics[width=0.5\textwidth]{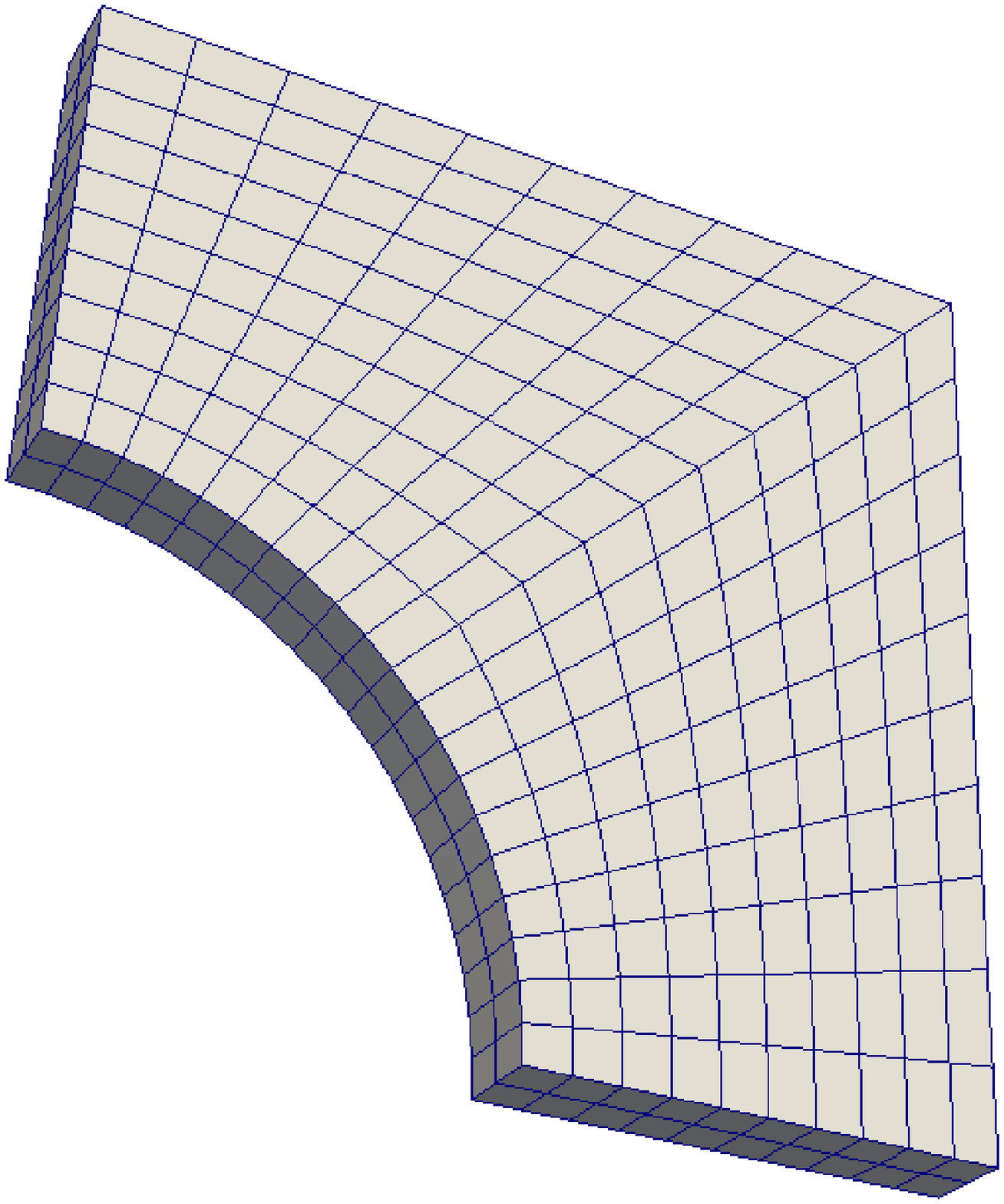}
  \includegraphics[width=0.5\textwidth]{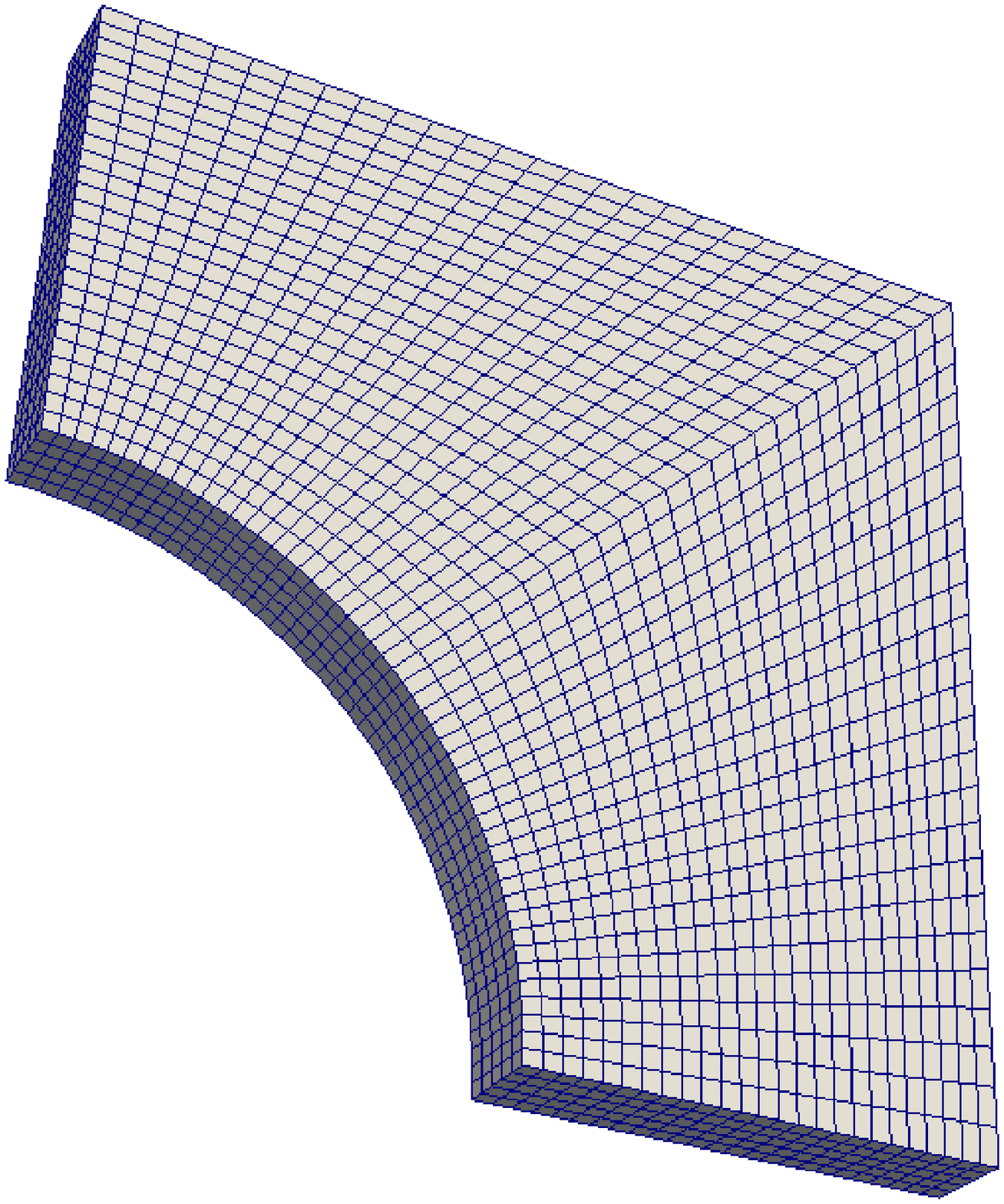}
}
 \end{picture}
 (b)
\end{minipage}
% \begin{minipage}{0.56\textwidth}
% \begin{tabular}{lll} \toprule
% % \multicolumn{3}{c}{\textbf{Material Data}} \\
% Parameter        & Value                                & Unit    \\ \midrule    %  & Description                    
% $E $                  &    $500$                        &      MPa           \\  %             & Elasticity modulus  
% $\nu$                  &   $0.3$                         &     -   \\  %                 & Poisson ratio   
% $r$                    &   $5.0$                          & MPa                  \\  %             &   Dissipation parameter 
% $\beta$               &    $\in\{10,100,1000\}$            & MPa$\,$mm$^2$    \\  %      &     Nonlocal parameter       
% $D_{\max}$       &    $0.95$                              & -                 \\       %             &   Maximal damage value    
% $d\hat{u}$            &   $\in\{0.25,0.1,0.025,0.01\}$   & mm    \\ \bottomrule
% % Element: & P2-P1-P0 & P2-P1B-P0 & P2-P1B-P1 & P2-P2-P1 & Q2-Q1-P0 & Q2-Q2-Q1 \\ \midrule
% \end{tabular}
% \vfill
% %  \includegraphics[draft,width=\textwidth]{fig/04_cm_nel2_mesh.pdf}
%  \begin{picture}(0,0)
% \put(-0.2,-0.8){(b)}
%  \end{picture}
% \end{minipage}
% \end{center}
\caption{(a) Description of the geometry of the plate with hole benchmark problem. 
(b) Example finite element meshes (400 and 6250 elements).}
\label{f:pwh_description}
\end{figure}

% pwh bvp description
As first example, we consider a plate with a circular hole which is subjected to the prescribed displacement ${\bfu}^\star=(0,{u}^\star,0)^T$ with ${u}^\star=25\, \text{mm}$ at its upper boundary ($Y=L$, with $L=100\, \text{mm}$).
The geometry of the considered domain is depicted in Fig.~\ref{f:pwh_description}(a).
Due to the symmetry of the problem, the computational domain amounts only to the upper right quarter of the geometry.
Moreover, as a consequence of the symmetry, the displacement is fixed in X-direction with $u_X=0$ at the left edge $X=0$ and in Y-direction with $u_Y=0$ at the lower edge $Y=0$. 
The radius of the circular hole as shown in Fig.~\ref{f:pwh_description}(a) is given with $R=50\, \text{mm}$ and the thickness is given with $H=10\, \text{mm}$.
An overview of the used material and load parameters is collected in Tab.~ \ref{t:parameters}.
\subsubsection{Convergence study}
% pwh convergence study
\begin{figure}
\unitlength1cm
\begin{center}
\begin{minipage}{0.48\textwidth}
 \includegraphics[width=\textwidth]{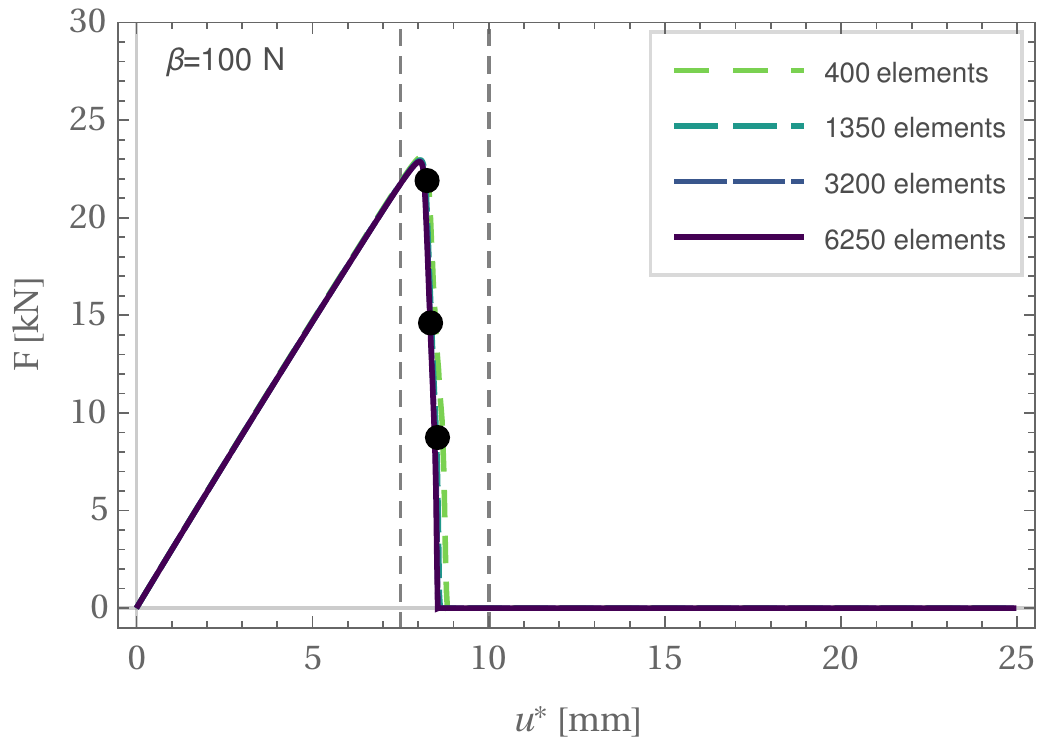}
 (a)
\end{minipage}
\begin{minipage}{0.48\textwidth}
 \includegraphics[width=\textwidth]{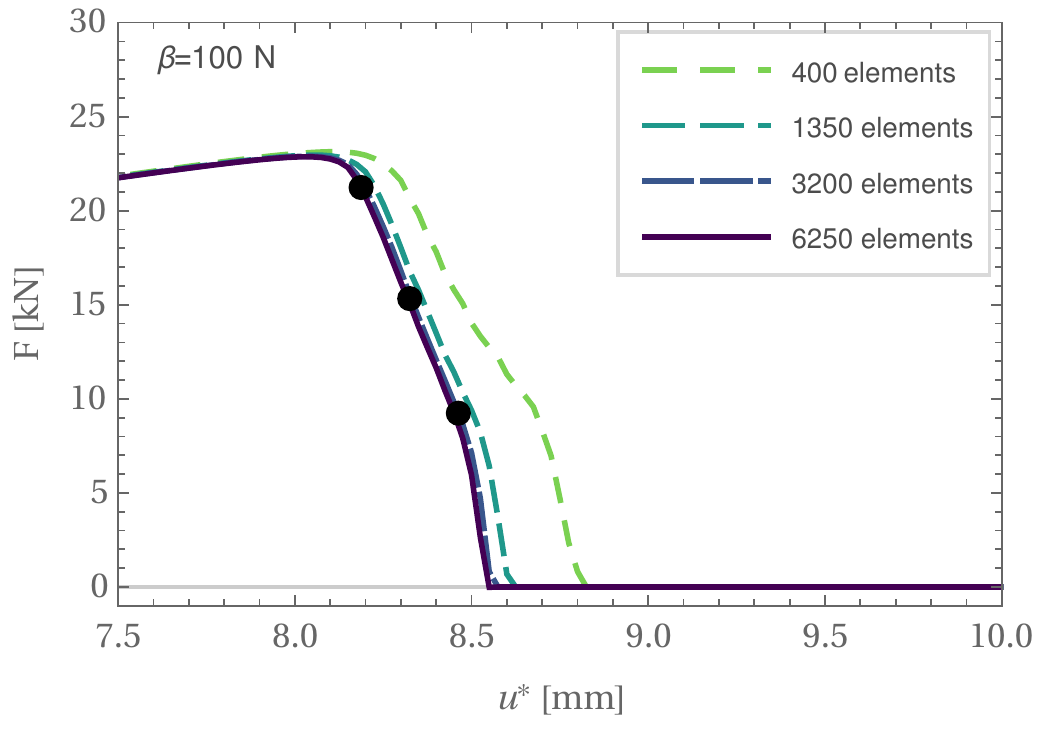}
 (b)
\end{minipage}
%
%contourplots showing damage evolution
\begin{picture}(12,6.8)
\put(-2,3.3){
\begin{minipage}{0.30\textwidth}
 \includegraphics[width=\textwidth]{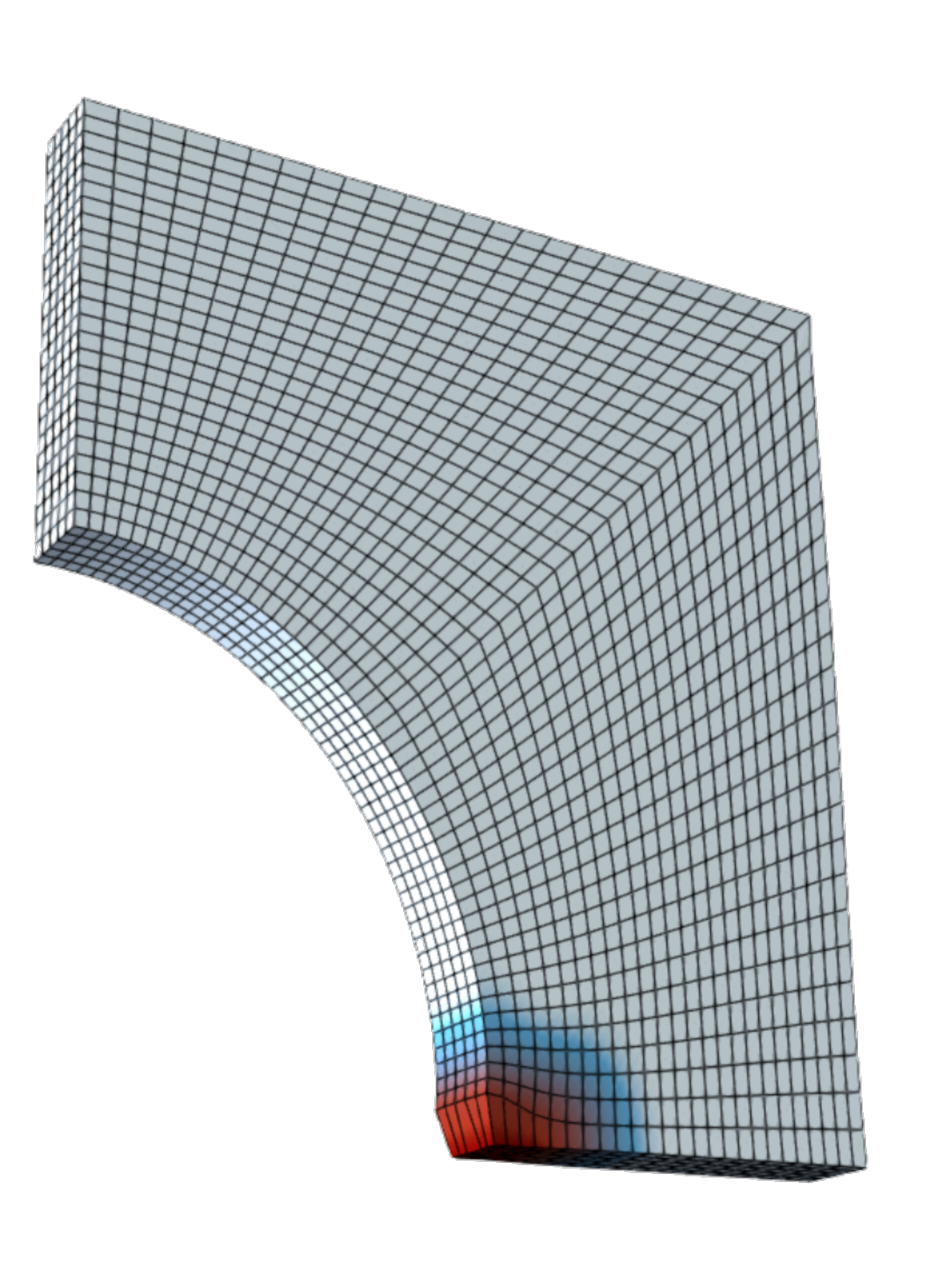}
\end{minipage}
\hspace{-1cm}
\begin{minipage}{0.30\textwidth}
 \includegraphics[width=\textwidth]{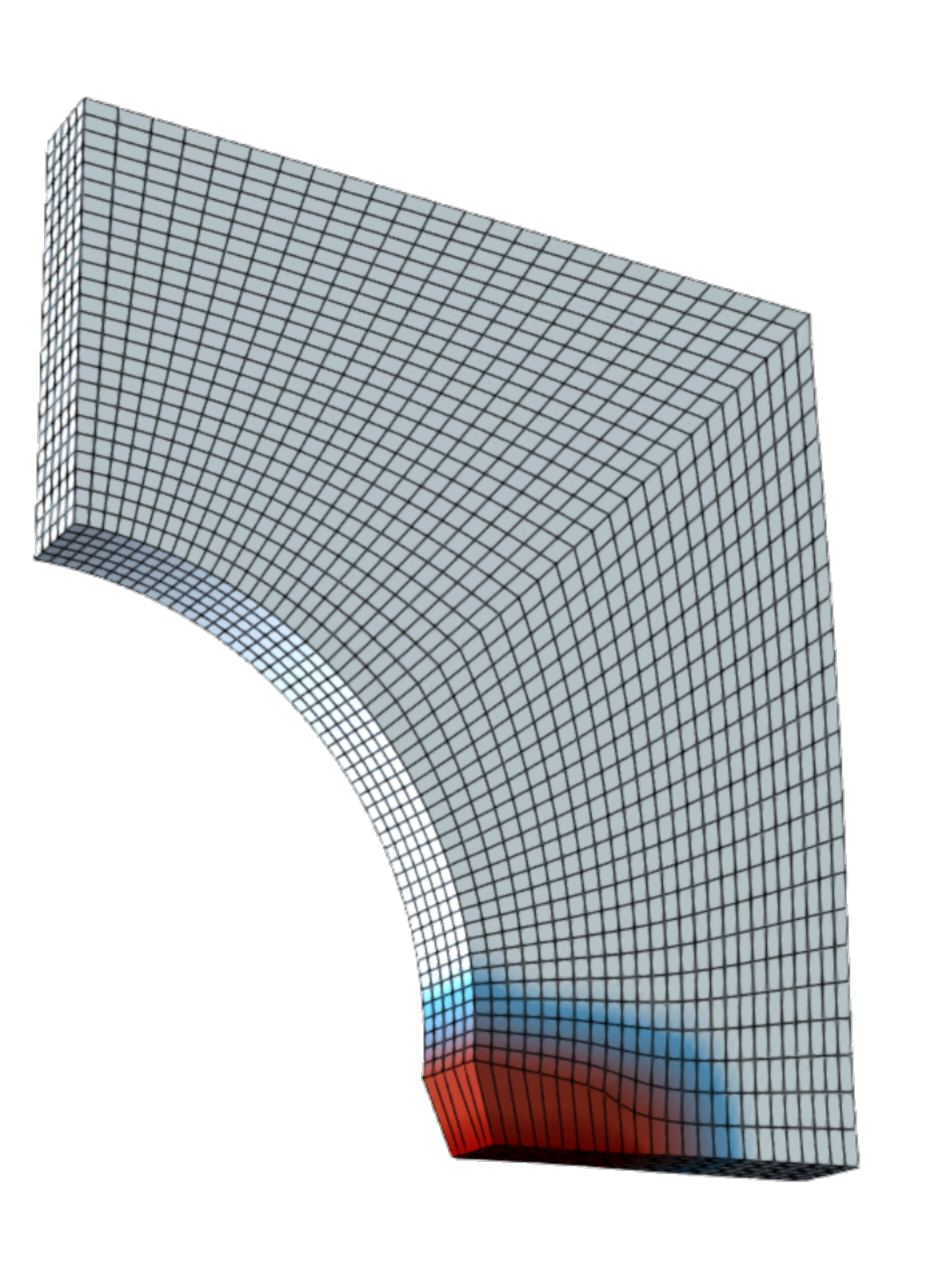}
%  (b)
\end{minipage}
\hspace{-1cm}
\begin{minipage}{0.30\textwidth}
 \includegraphics[width=\textwidth]{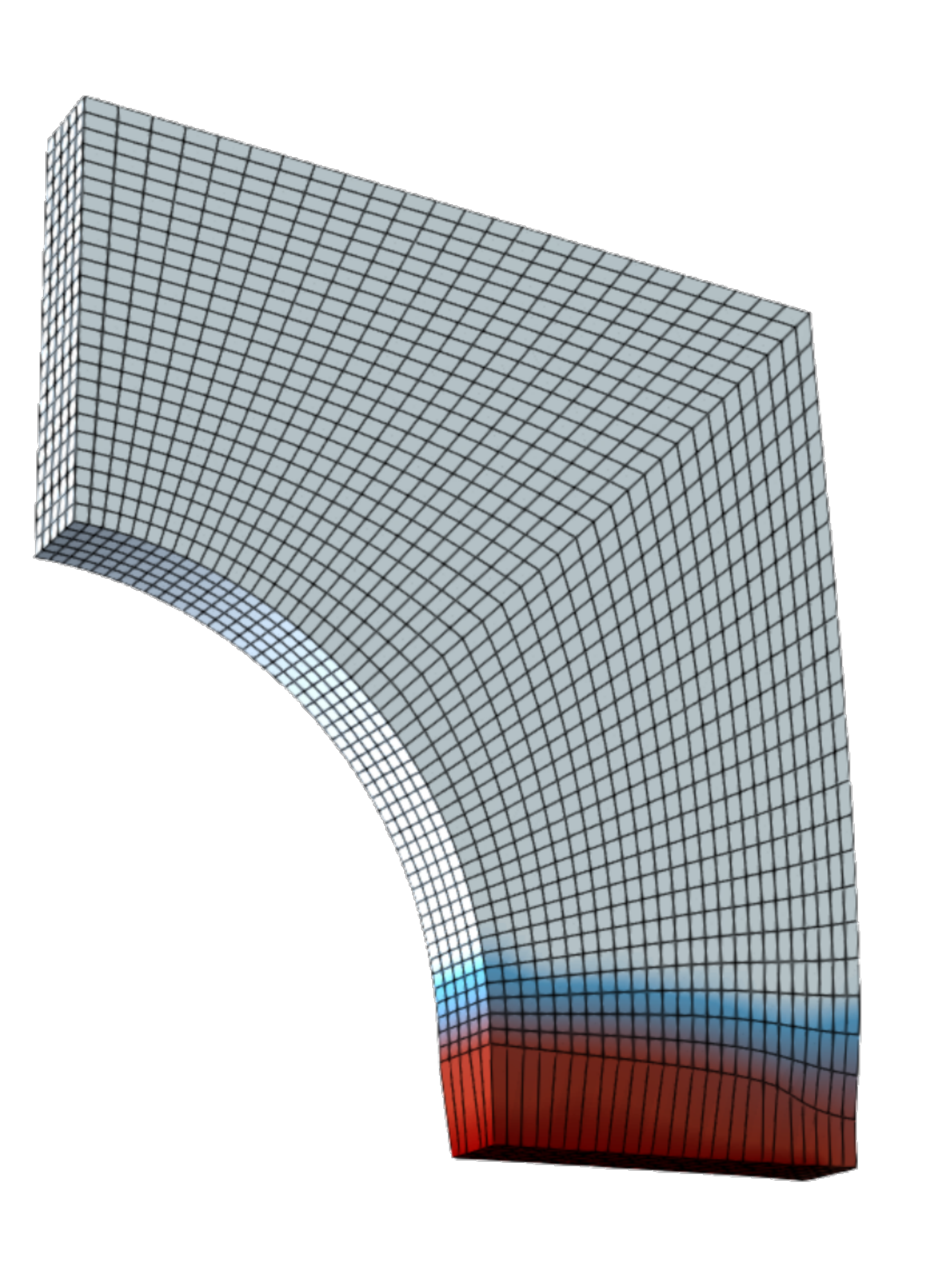}
%  (c)
 \begin{picture}(0,0)
\put(5.5,1.2){
 \includegraphics[width=1.4cm]{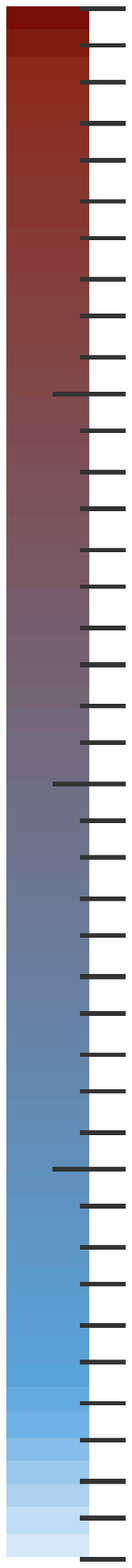}
}
\put(5.6,6.9){{\small $D$}}
\put(6.15,1.1){{\small $0.00$}}
\put(6.15,3.85){{\small $0.46$}}
\put(6.15,6.6){{\small $0.95=D_\text{crit}$}}
 \end{picture}
\end{minipage}
}
\put(-2,0){(c)}
\end{picture}
\end{center}
\caption{(a) Force/displacement curves (higher resolution in (b)) and (c) corresponding contourplots depicting the damage evolution for the 6250 element mesh in three different load steps which are depicted as bullets in (a).}
\label{f:pwh_beta_1e2_f_u_and_contourplots}
\end{figure}

% pwh convergence study
In Fig.~\ref{f:pwh_beta_1e2_f_u_and_contourplots}(a), force/displacement curves corresponding to various mesh refinements (see eg. Fig.~\ref{f:pwh_description}(b)) are shown.
The value of the prescribed displacement ${u}^\star$ is depicted on the abscissa whereas the value of the reaction force $F$ in Y-direction, which is recovered from the nodes at the upper surface $Y=L$, is plotted on the ordinate.
Here, the nonlocal parameter has the value $\beta=100\,\text{N}$ and the prescribed displacement is applied in increments of $0.025$ mm, which is equivalent to $1000$ load steps.
For the given plot resolution in Fig.~\ref{f:pwh_beta_1e2_f_u_and_contourplots}(a), it can be observed that the force/displacement curve of the lowest refinement stage (400 elements) almost coincides with the curves corresponding to the higher refinement stages.
To illustrate the convergence of the curves as the number of elements increases, an enlarged resolution of the plot is given in Fig.~\ref{f:pwh_beta_1e2_f_u_and_contourplots}(b).
The area of the enlargement is the transition regime at which material softening occurs (marked with gray dashed lines in Fig.~\ref{f:pwh_beta_1e2_f_u_and_contourplots}(a)).
The evolution of the material softening is also illustrated through the contourplots of the damage function $D$ in Fig.~\ref{f:pwh_beta_1e2_f_u_and_contourplots}(c).
The contourplots correspond to the computation with the 6250 element mesh and the depicted stages of damage evolution are marked with bullets in Fig.~\ref{f:pwh_beta_1e2_f_u_and_contourplots}(a) and (b).
It can be observed that already at low damage propagation states, as depicted in the left contourplot, the damage value $D$ of several elements reaches the critical damage value $D_{\mathrm{crit}}$.
For all computations, convergence of the iterative solution procedure was obtained.

\subsubsection{Influence of the nonlocal parameter}
% pwh varbeta 
\begin{figure}
\unitlength1cm
\begin{center}
%
%
% beta=10
\begin{minipage}{0.48\textwidth}
\begin{picture}(0,0)
\put(0,0.5){(a)}
 \end{picture}
\includegraphics[width=\textwidth]{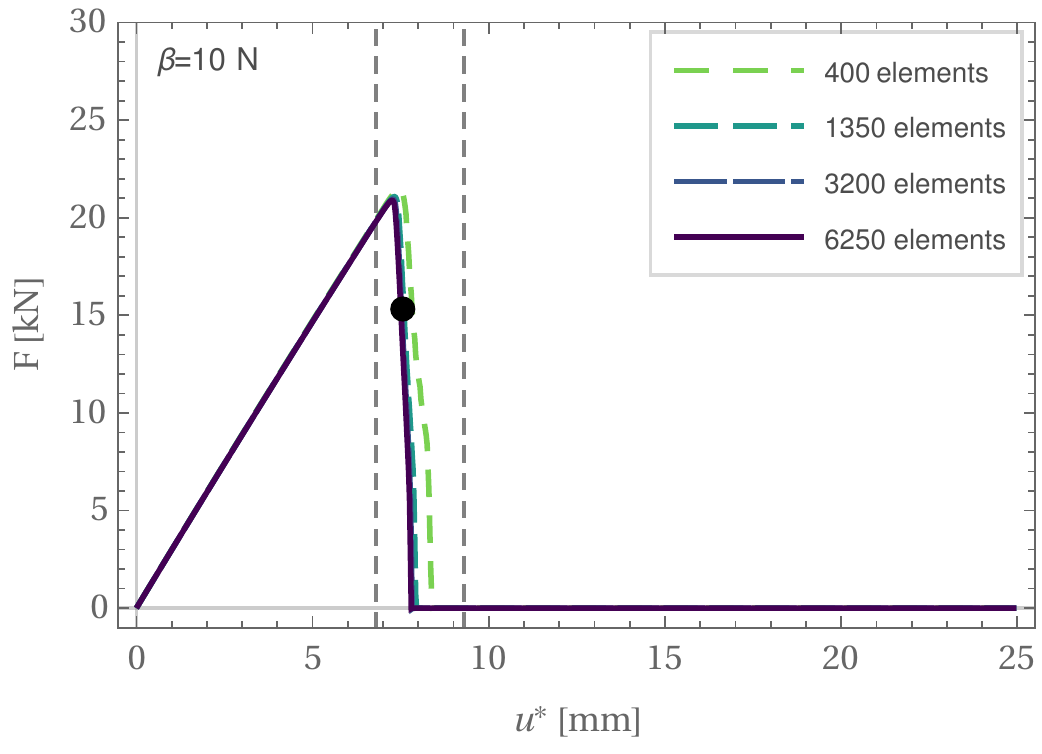}
\end{minipage}
\begin{minipage}{0.48\textwidth}
 \includegraphics[width=\textwidth]{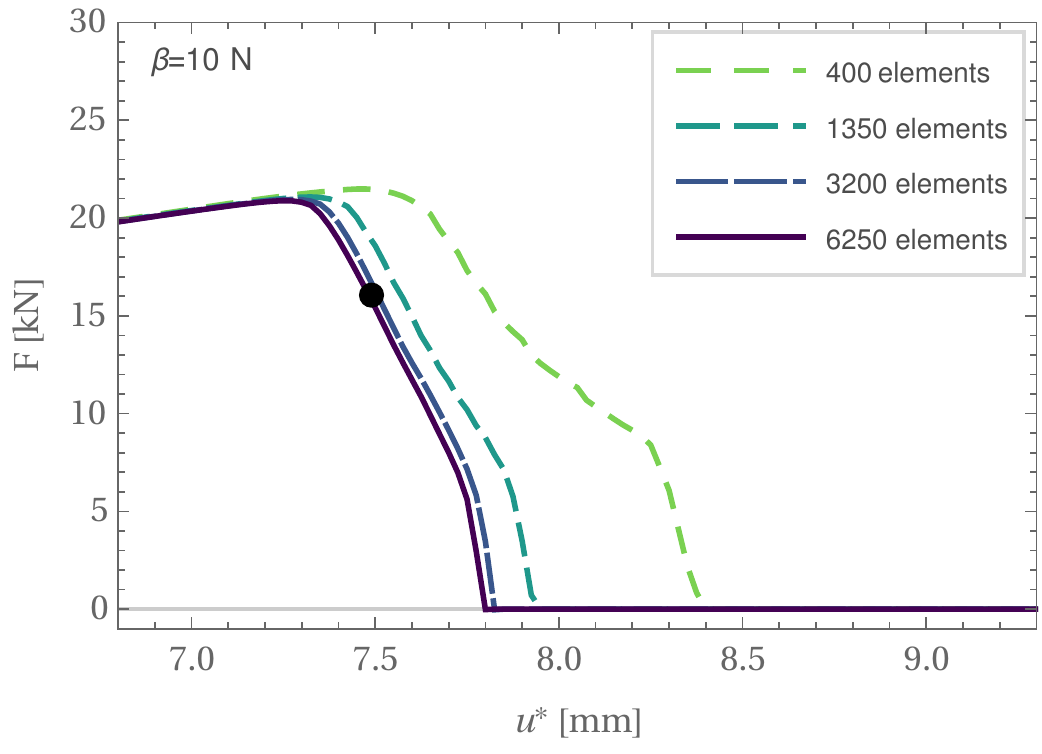}
\end{minipage}
%
% beta=100
\begin{minipage}{0.48\textwidth}
\begin{picture}(0,0)
\put(0,0.5){(b)}
 \end{picture}
 \includegraphics[width=\textwidth]{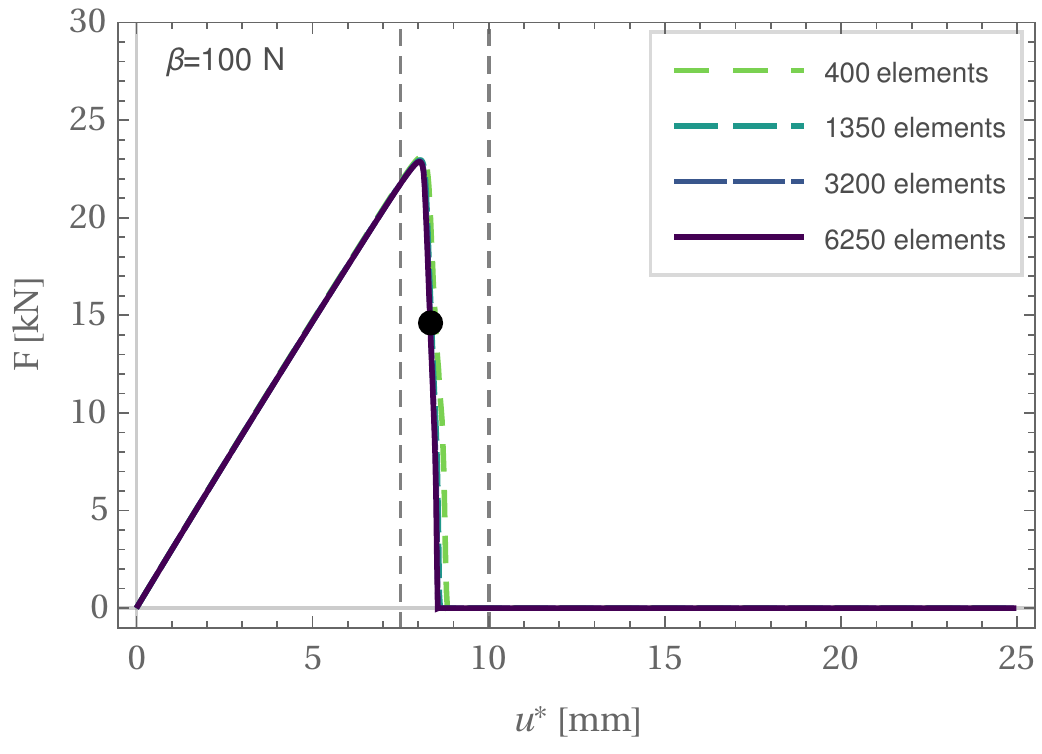}
\end{minipage}
\begin{minipage}{0.48\textwidth}
 \includegraphics[width=\textwidth]{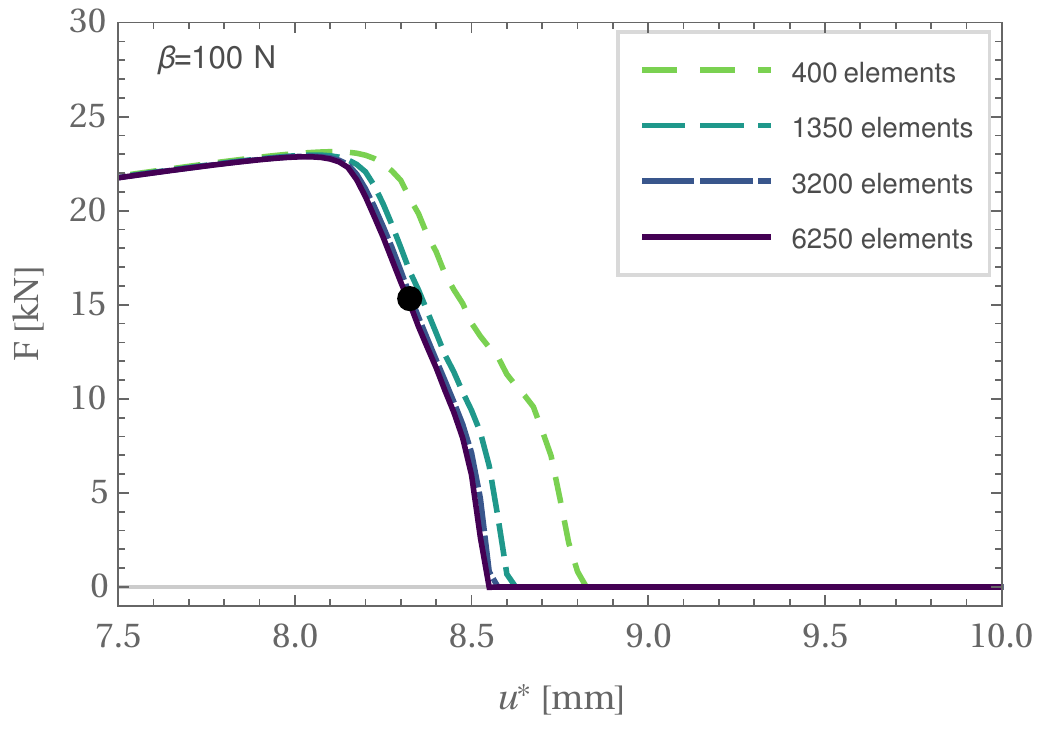}
\end{minipage}
%
% beta=1000
\begin{minipage}{0.48\textwidth}
\begin{picture}(0,0)
\put(0,0.5){(c)}
 \end{picture}
 \includegraphics[width=\textwidth]{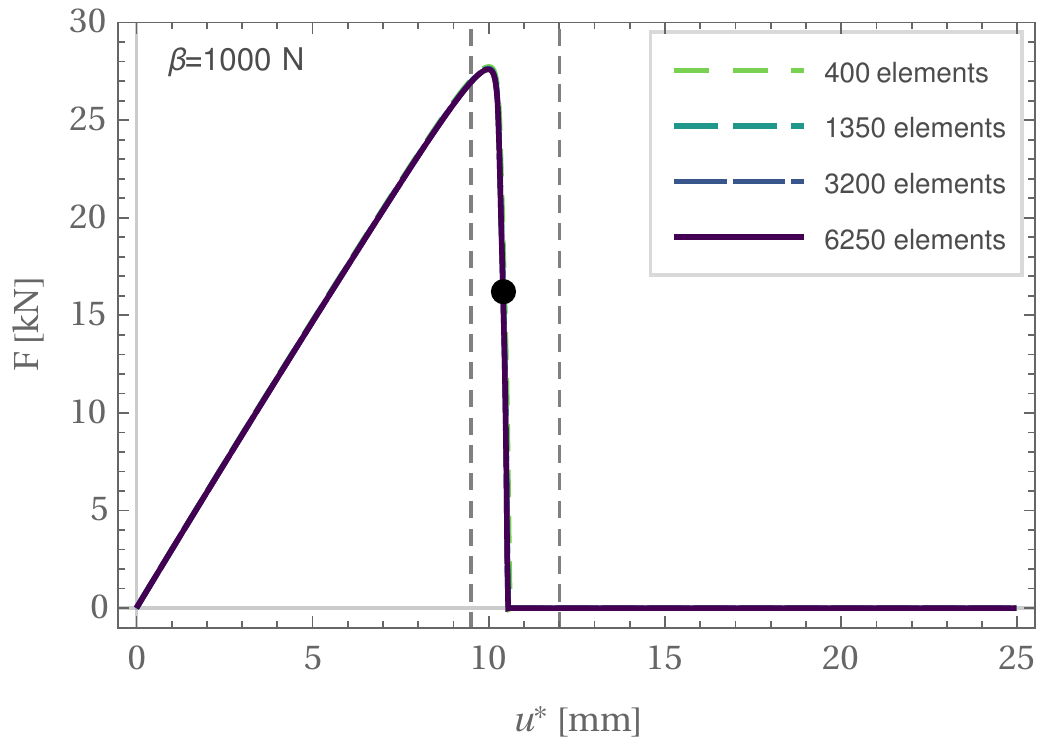}
\end{minipage}
\begin{minipage}{0.48\textwidth}
 \includegraphics[width=\textwidth]{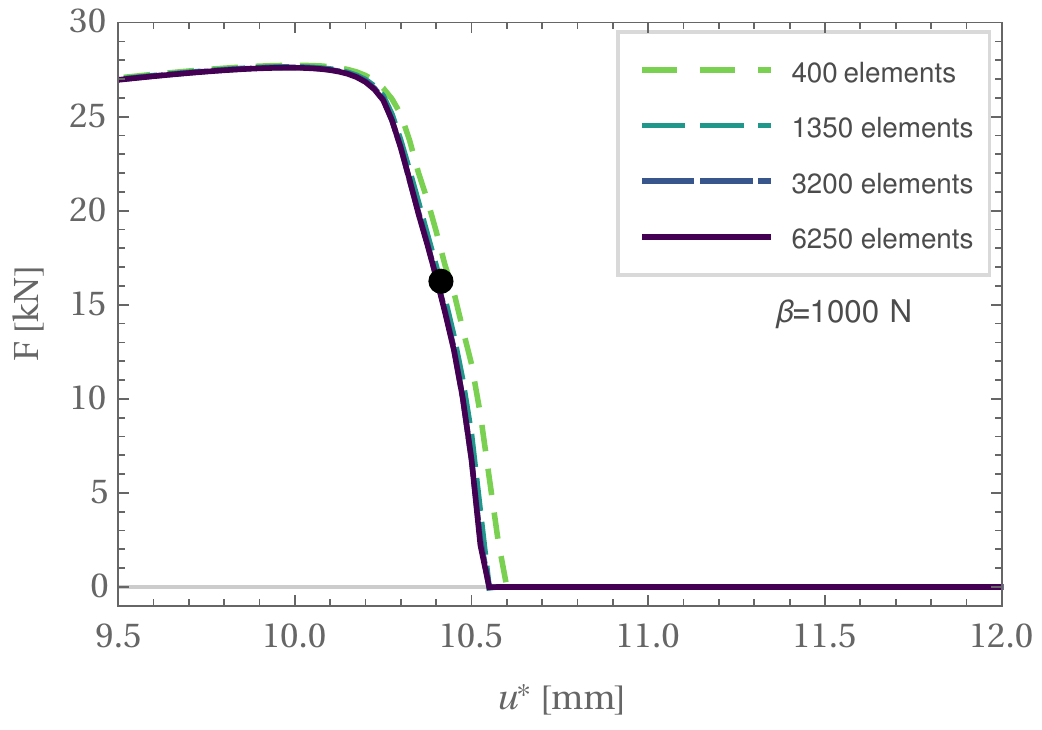}
\end{minipage}
\end{center}
\caption{Influence of the nonlocal parameter: force/displacement curves for nonlocal parameters  with enlarged resolution on the right hand side.
Figures (a), (b) and (c) correspond to parameters $\beta=\{10,100,1000\}\,\text{N}$, respectively.}
\label{f:pwh_F_over_u_varbeta}
\end{figure}

% influence of the nonlocal parameter contourplotsp
\begin{figure}
\unitlength1cm
\begin{center}
\begin{picture}(12,6.8)
\put(-2.1,3.3){
\begin{minipage}{0.30\textwidth}
 \includegraphics[width=\textwidth]{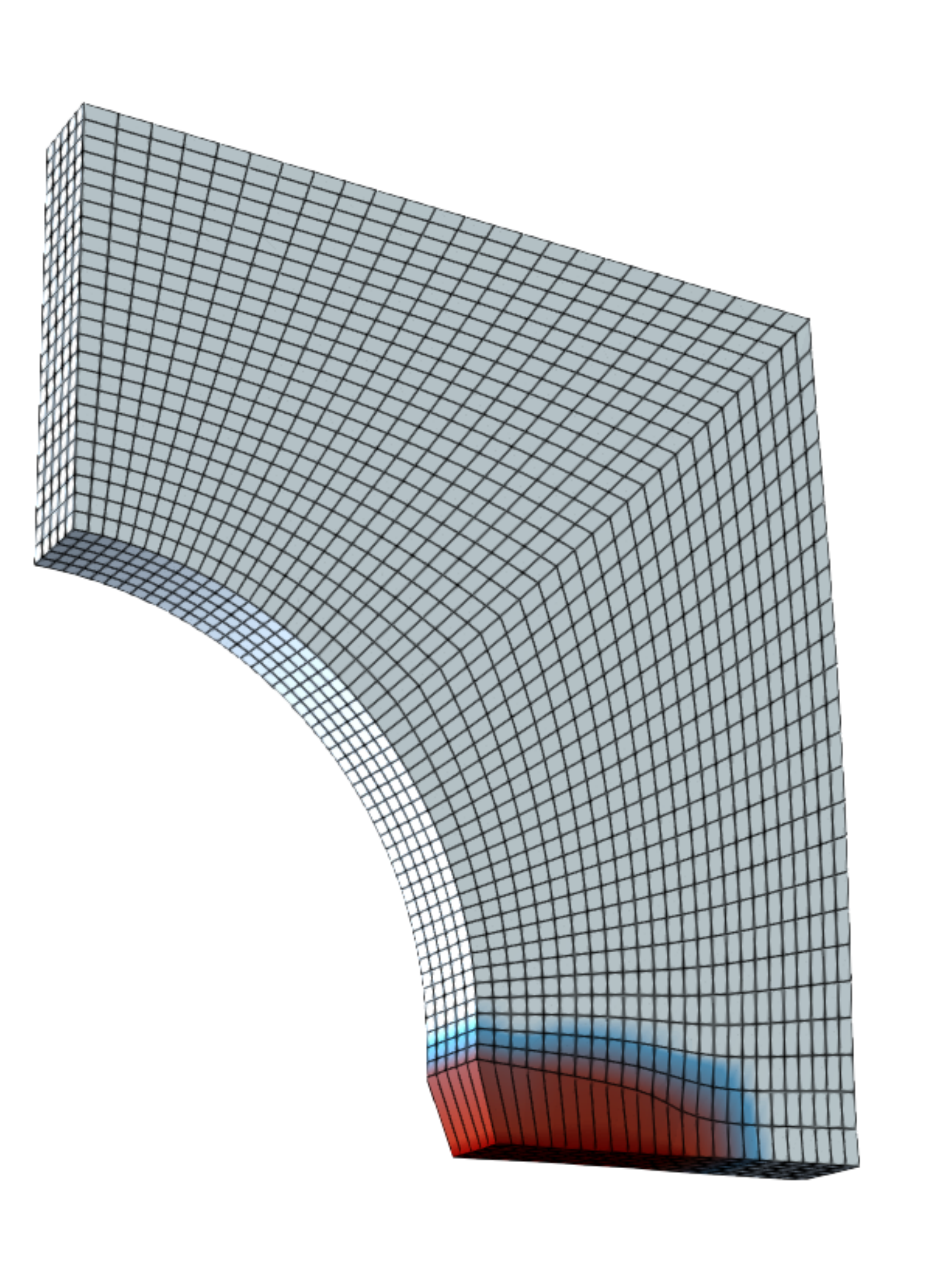}
\end{minipage}
 \hspace{-0.8cm}
\begin{minipage}{0.30\textwidth}
 \includegraphics[width=\textwidth]{fig/pwh_r4_beta_1e2_alpha_5e-2_332_of_1000.pdf}
%   (b)
\end{minipage}
 \hspace{-0.8cm}
\begin{minipage}{0.30\textwidth}
 \includegraphics[width=\textwidth]{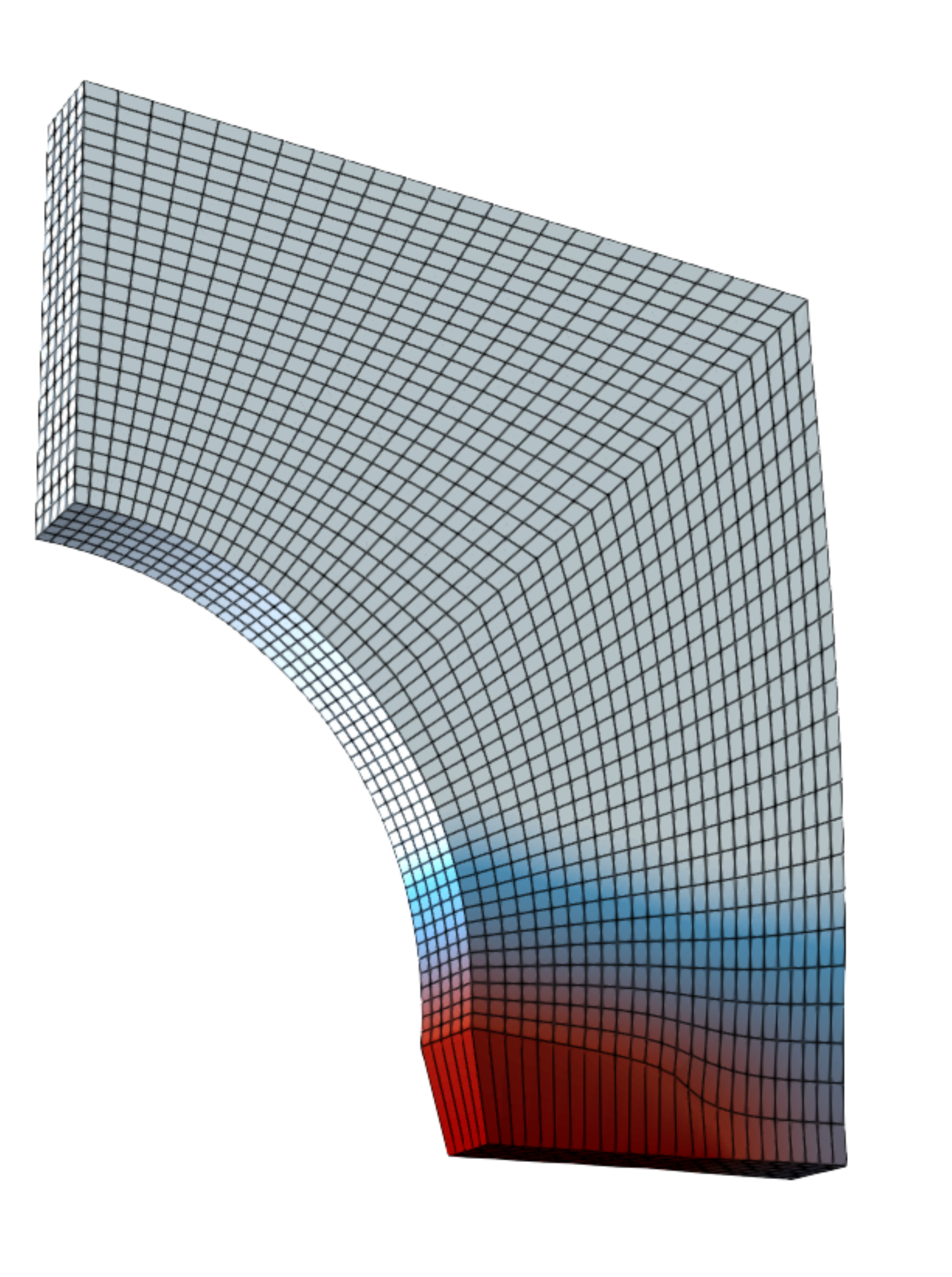}
%  (c)
 \begin{picture}(0,0)
\put(5.1,1.){
 \includegraphics[width=1.4cm]{fig/d_scale_alpha_5e-2.eps}
}
\put(5.2,6.7){{\small $D$}}
\put(5.75,0.9){{\small $0.00$}}
\put(5.75,3.65){{\small $0.46$}}
\put(5.75,6.4){{\small $0.95=D_\text{crit}$}}
 \end{picture}
\end{minipage}
}
% \put(-2,0){(c)}
\end{picture}
\end{center}
\caption{Influence of the nonlocal parameter: contourplots for varying values of $\beta$ (from left to right) corresponding to the bullet marks in Fig.~\ref{f:pwh_F_over_u_varbeta}(a)-(c).
Left: $\beta=10\,\text{N}$, middle: $\beta=100\,\text{N}$, right: $\beta=1000\,\text{N}$.
% (left: red, middle: black, right: blue bullet)
}
\label{f:pwh_contourplots_varbeta}
\end{figure}

% influence of the nonlocal parameter description
In this subsection, the influence of the value of the nonlocal parameter $\beta$ on the convergence behavior and the damage evolution is investigated.
Therefore, in Fig.~\ref{f:pwh_F_over_u_varbeta} corresponding force/dis\-place\-ment curves are shown for values $\beta=10,100$ and $1000\,\text{N}$, respectively.
The same load increment of $0.025\,\text{mm}$ as in the previous subsection is used.
From the plots of Fig.~\ref{f:pwh_F_over_u_varbeta}, it can be observed that as $\beta$ increases %and thus, the degree of regularization increases, 
the convergence of the curves appears more rapidly.
However, as $\beta$ increases also the starting point of damage initialization is shifted to higher load values. 
This goes along with the increased nonlocality of the distribution of the damage function $D$ as depicted in the contourplots of Fig.~\ref{f:pwh_contourplots_varbeta}(c):
while in the left contourplot ($\beta=10\,\text{N}$) the damage distribution is mostly concentrated to the elements adjacent to the lower edge, in the right contourplot ($\beta=1000\,\text{N}$), a more smeared damage distribution can be observed.
% For all values of $\beta$ convergence of the iterative solution procedure was obtained.%, even for rather small values, 
% the solution procedure is observed to behave with similar robustness.% as previously discussed.

% % influence of the nonlocal parameter F over nelem
% \begin{figure}
% % \setlength\textwidth{12.0cm}
% \unitlength1cm
% \begin{center}
% \begin{minipage}{0.48\textwidth}
%  \includegraphics[width=\textwidth,height=5.5cm]{fig/pwh_F_over_nelem.pdf}
% \end{minipage}
% \end{center}
% \caption{Influence of the nonlocal parameter: convergence of the normed reaction forces for varying nonlocal parameters. 
% The values correspond to the load stages marked with bullets in figure \ref{f:pwh_F_over_u_varbeta}}
% \label{f:pwh_F_over_nelem}
% \end{figure}

\subsubsection{Influence of the load increment and computing efficiency}

\begin{figure}
 \unitlength1cm
 \begin{minipage}{0.48\textwidth}
 \includegraphics[width=\textwidth]{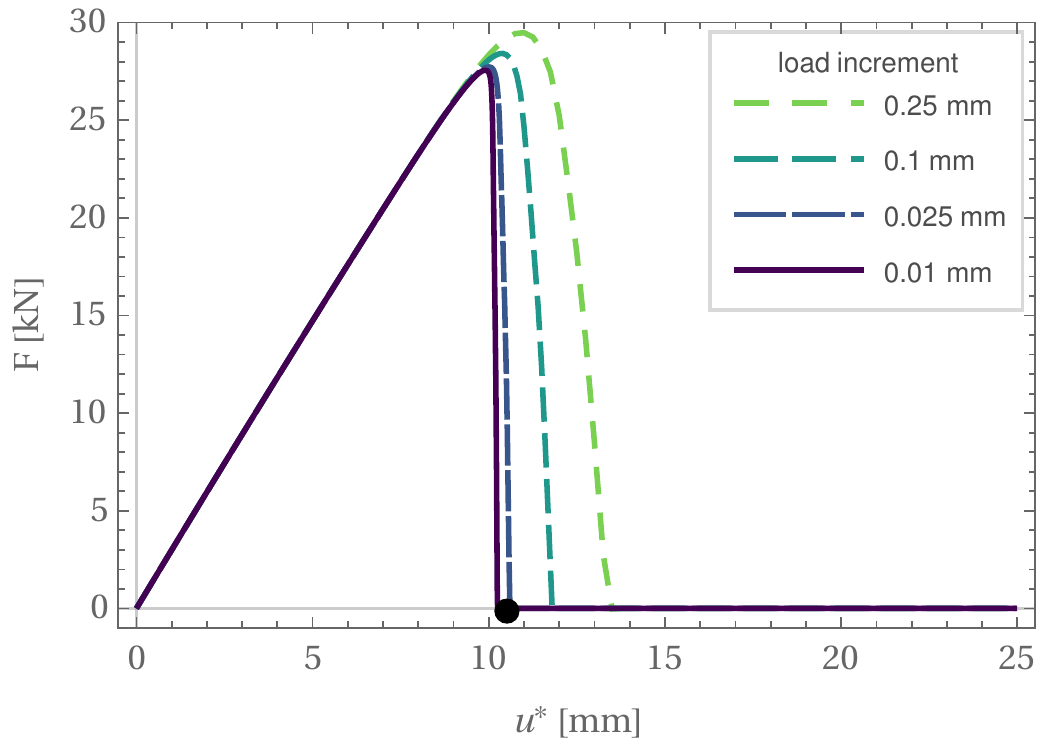}
 (a)
\end{minipage}
\begin{minipage}{0.48\textwidth}
 \includegraphics[width=\textwidth]{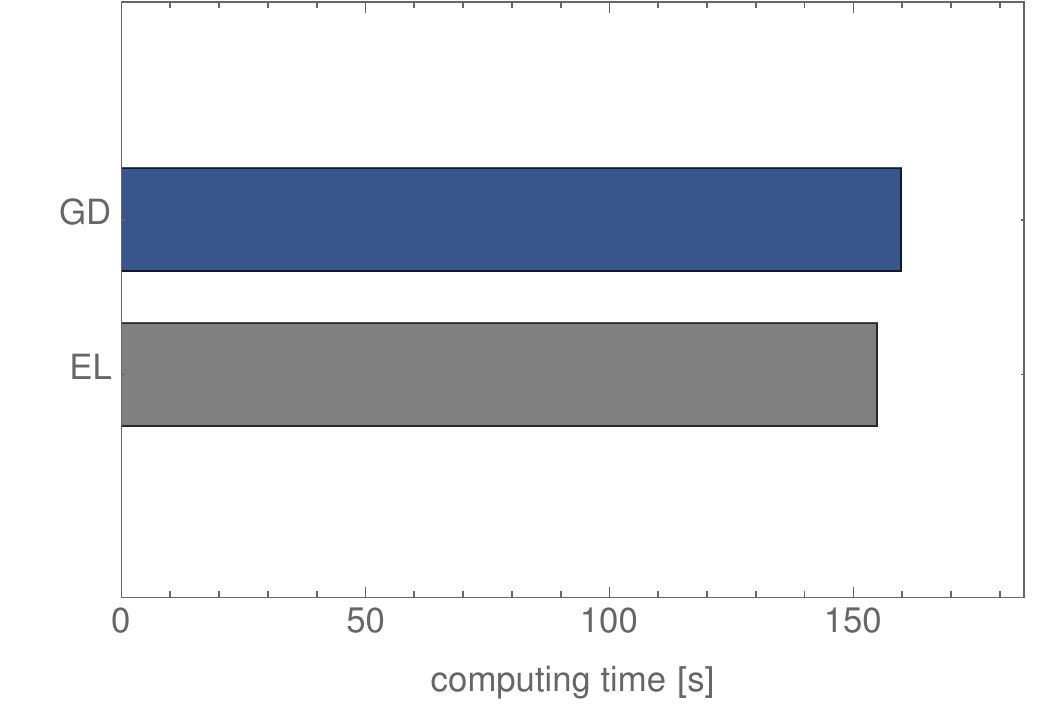}
 (b)
\end{minipage}
 \caption{
(a) Influence of the load increment on the solution: 
force/displacement curves for varying load increments for a 400 element mesh and $\beta=1000\, \text{N}$.
 (b) GD: computing time until reaching the marked point at $u^\star=10.85\,\text{mm}$ in (a) with the load increment $0.025\,\text{mm}$. 
 EL: computing time of a purely elastic 8-node hexahedral reference element until the same load of $u^\star=10.85\,\text{mm}$.}
\label{f:pwh_var_load_increment}
\end{figure}

To investigate the influence of the load increment on the solution, we show the convergence of the force/displacement curves for the specific displacement increments $\{0.25,0.1,0.025,0.01\}\, \text{mm}$,
while the mesh refinement stage (400 elements) and nonlocal parameter ($\beta=1000\, \text{N}$) remain fixed.
For the chosen parameters, the corresponding curves are depicted in Fig.~\ref{f:pwh_var_load_increment}(a) and show the tendency to converge as the value of the load increment decreases and, consequently, the number of steps increases.
% computing efficiency
Furthermore, Fig.~\ref{f:pwh_var_load_increment}(b) shows the computing time for the gradient damage model (labeled with GD) and a comparative purely elastic simulation (labeled with EL). The time is measured until the crack has completely evolved which corresponds to a load of $u^\star=10.85\,\text{mm}$. Considering also later loading steps would result in an overestimation of the computation time of the elastic simulation: whereas for the elastic material behavior a non-linear finite element simulation has to be performed which requires more than two iteration steps for convergence, the damage model has to solve the rigid body movement due to the completely evolved horizontal crack that separated the plate. Consequently, only two iteration steps are required for convergence yielding a total computing time that is larger for the elastic problem than for the damage model. The elastic problem has been solved by deactivating the update subroutine for the damage field. This implies that the same effort for writing the results to text files was needed and no unfair acceleration of either simulation has been employed. For reaching the critical load of $u^\star=10.85\,\text{mm}$, the elastic simulation required $t=154.9\,$s while the damage simulation finished after $t=159.8\,$s. This computing time increase of $3.2\, \%$ of the proposed approach compared to the purely elastic reference simulation unveils that the additional time corresponding to the damage update represents only a small portion of the overall computing time. For instance, if instead a full finite element scheme for the balance of linear momentum as well as for the microstructure evolution leading to four nodal unknowns is employed, the computation costs time can easily exceed 10 times the computing time required for the purely hyperelastic problem. A comparable observation has also been made for the evaluation of the damage model in the context of small deformations in \cite{junker2019fast}.

\subsection{U-shaped geometry}

% u-shape plots
\begin{figure}
\unitlength1cm
\begin{center}
\begin{minipage}{0.48\textwidth}
\begin{picture}(7.5,17)
\put(0,12.6){
 \def\svgwidth{0.9\textwidth}
  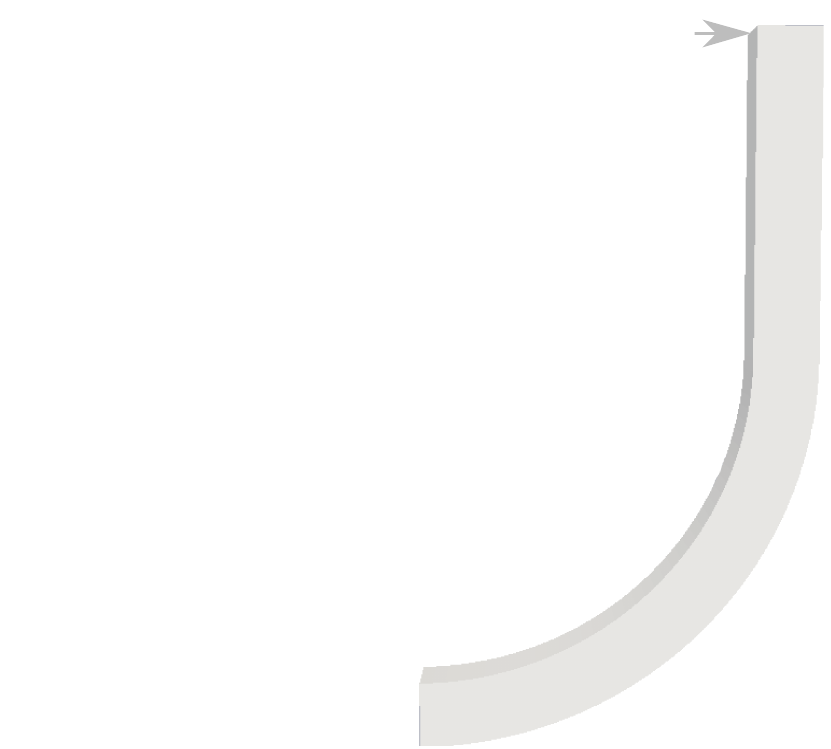}
\put(0,12.6){(a)}
\put(0,5.5){\includegraphics[width=\textwidth]{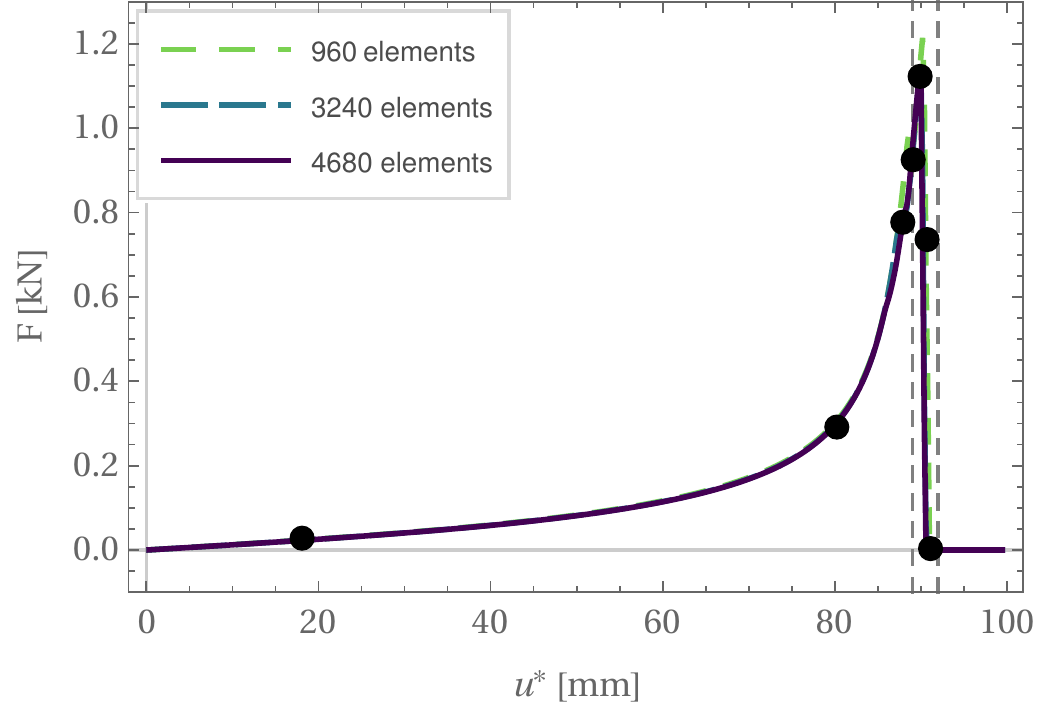}}
\put(0,0){\includegraphics[width=\textwidth]{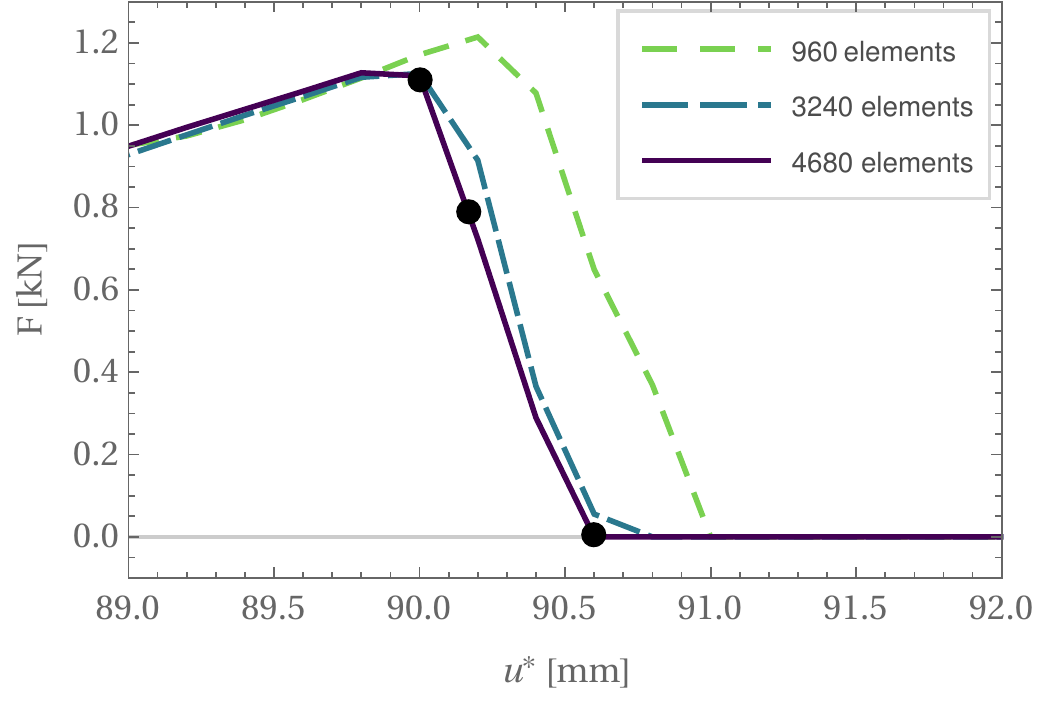}}
% upper plot
\put(2.3,7.1){{\small $1$}}
\put(6.3,7.9){{\small $2$}}
\put(6.8,9.35){{\small $3$}}
\put(6.9,9.9){{\small $4$}}
\put(7.6,10.5){{\small $5$}}
\put(7.6,9.2){{\small $6$}}
\put(7.00,6.8){{\small $7$}}
% lower plot
\put(3.0,4.5){{\small $5$}}
\put(3.3,3.75){{\small $6$}}
\put(4.3,1.35){{\small $7$}}
\put(0,0){(b)}
\end{picture}
\end{minipage}
\begin{minipage}{0.48\textwidth}
 \begin{picture}(7.5,17)
\put(0,13){\includegraphics[scale=0.28]{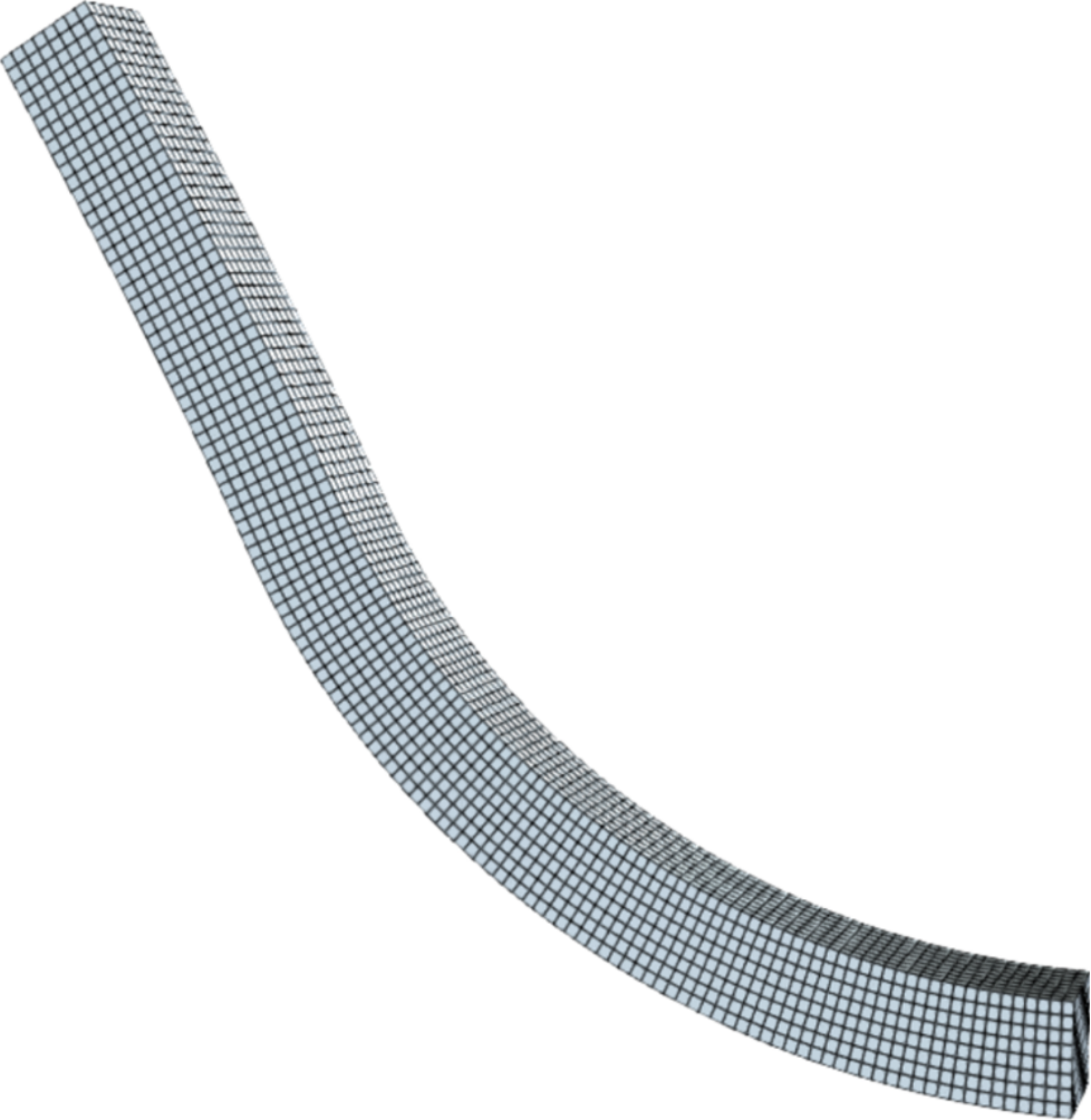}}
\put(7,13){\includegraphics[scale=0.28]{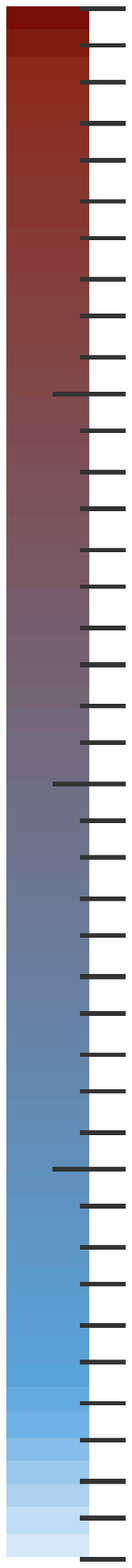}}
\put(6.95,18.8){{\small$D$} }
\put(7.5,12.9){{\small $0.000$}}
\put(7.5,15.7){{\small $0.500$}}
\put(7.5,18.5){{\small $0.995=D_\text{crit}$}}

\put(0,11){\includegraphics[scale=0.28]{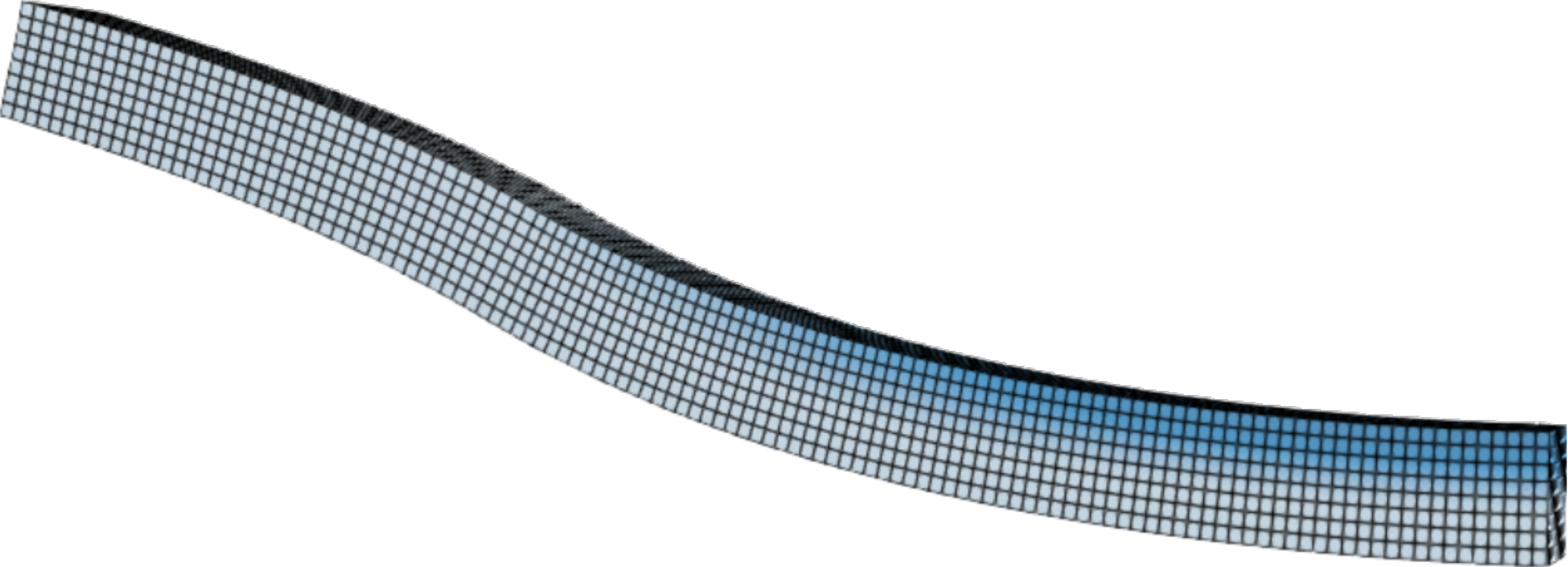}}
\put(0,9){\includegraphics[scale=0.28]{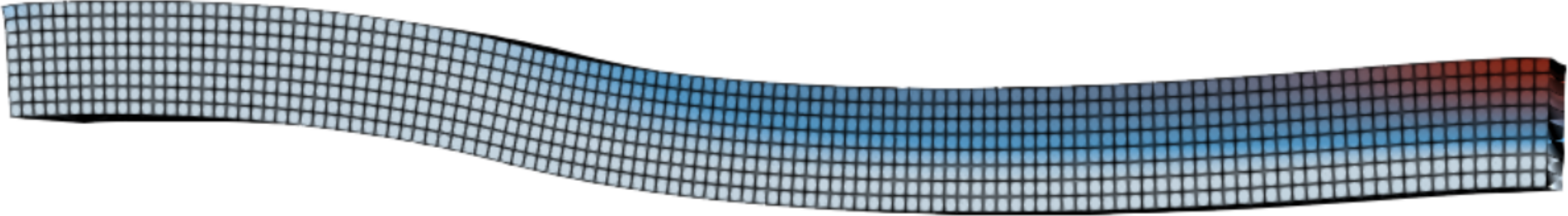}}
\put(0,7){\includegraphics[scale=0.28]{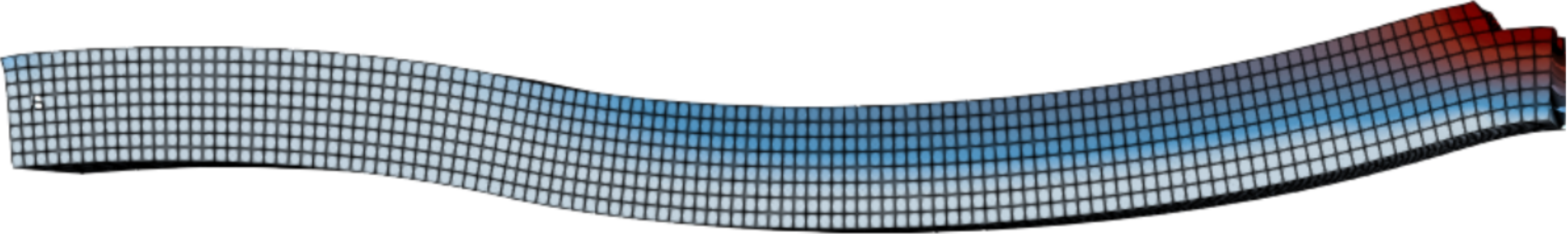}}
% \put(0,7){\includegraphics[scale=0.3]{fig/u_shape_r2_446_of_500.pdf}}
\put(0,5){\includegraphics[scale=0.28]{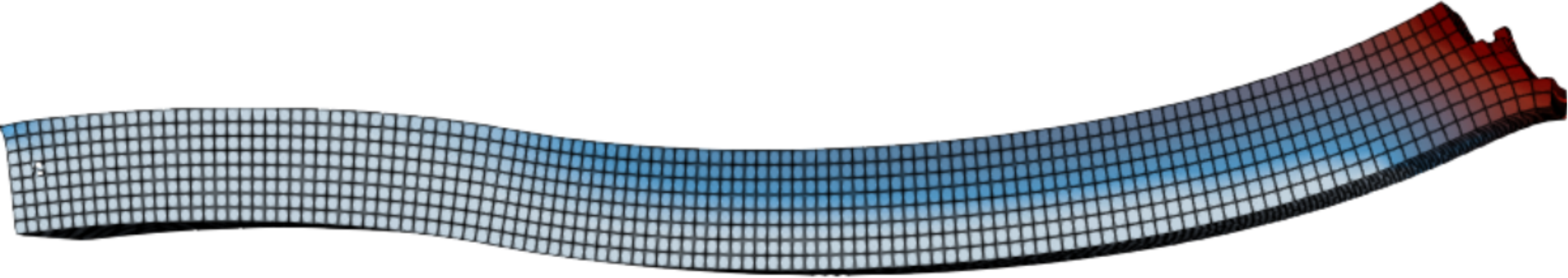}}
\put(0,3){\includegraphics[scale=0.3]{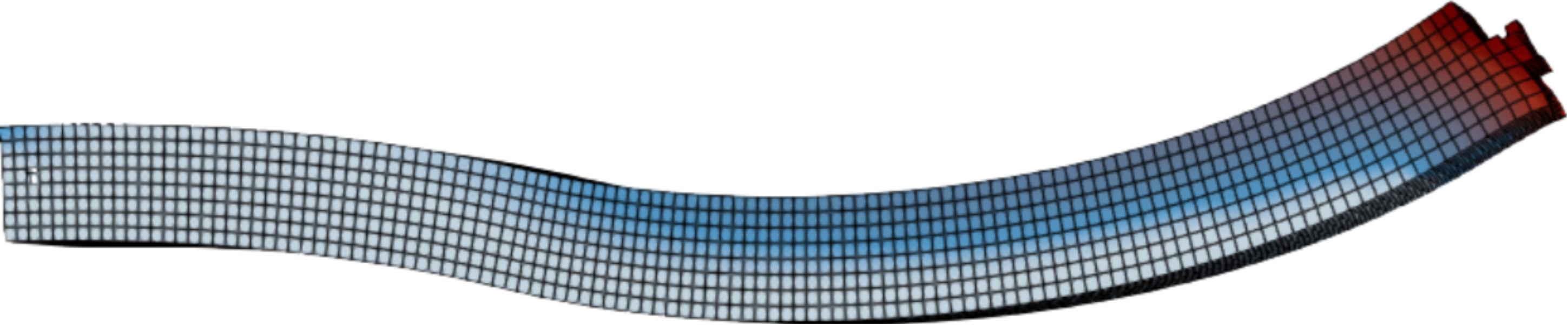}}
\put(0,0){\includegraphics[scale=0.28]{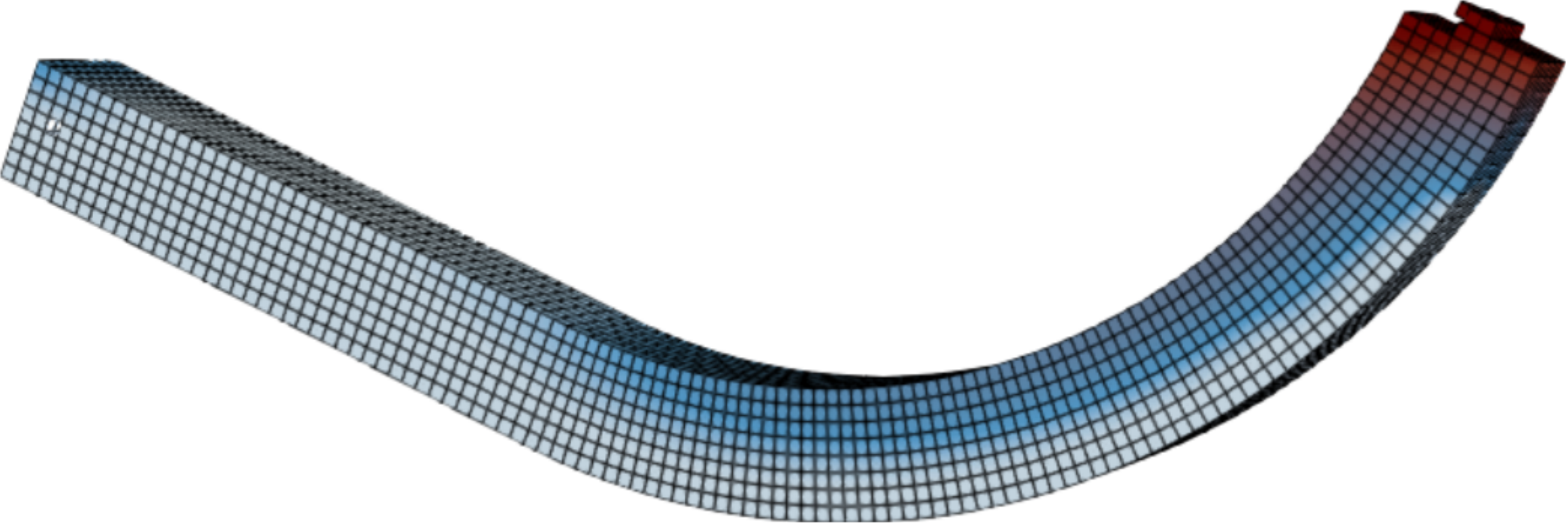}}
\put(0,14){{\small $1)$}}
\put(0,12){{\small $2)$}}
\put(0,9){{\small $3)$}}
\put(0,6.6){{\small $4)$}}
\put(0,4.6){{\small $5)$}}
\put(0,2.9){{\small $6)$}}
\put(0,0.8){{\small $7)$}}
\put(0,0){(c)}
 \end{picture}
\end{minipage}
% \put(-2,0){(c)}
%
%
\end{center}
\caption{(a) Description of the geometry of the u-shaped boundary value problem. 
(b) Force/displacement curves for various mesh refinement stages (enlarged resolution in the bottom). 
(c) Contourplots of the u-shaped problem visualizing the damage evolution corresponding to the marked load stages depicting the damage evolution.
}
\label{f:u_shape}
\end{figure}

% u-shape description
As second numerical test, a u-shaped geometry is analyzed in which, caused by its geometry and boundary conditions, large deformations are present (cf. Fig.~\ref{f:u_shape}(a)) although the strains may be comparatively moderate.
Due to its symmetry, only the left half of the domain is considered and the displacement $u_X=0$ is fixed at the symmetry plane $X=0$.
The dimensions (cf. Fig.~\ref{f:u_shape}(a)) are given with $R=50\,\text{mm}$ and $S=10\,\text{mm}$.
The prescribed displacement ${\bfu}^\star=({u}^\star,0,0)$ with ${u}^\star=100\,\text{mm}$ is imposed on the upper right edge $(X,Y)=(-50,50)\,\text{mm}$.
The used material and boundary parameters are shown in Tab.~\ref{t:parameters}.
From the force/displacement curves shown in the upper plot in Fig.~\ref{f:u_shape}(b) and with refined resolution in the lower figure, the tendency to converge can be concluded.
Due to the relatively large displacement, the additional material nonlinearity becomes visible in the force/displacement curves (pile-up of the reaction force close to failure).
The contourplots in Fig.~\ref{f:u_shape}(c) illustrate the damage propagation at the load stages marked with bullets in the force/displacement plots.
Here, to visualize the crack formation in the entirely damaged area of the domain, the elements in which the damage value has reached the upper bound $D^\eli=D_{\text{crit}}=0.995$ were made invisible. A remarkable elastic snapback is observed once the geometry fails, cf. Fig.~\ref{f:u_shape}(c)~7).
Despite the relatively coarse time discretization, convergence of the iterative solution was obtained at all refinement stages. It is worth mentioning that the snapback phenomenon constitutes a non-trivial process in a numerical treatment. Since we also observe convergence for this regime, the u-shaped boundary value problem empirically shows the numerical robustness of our approach to damage modeling for large deformations.

\section{Conclusions}
We presented a novel gradient damage model for hyperelastic structures undergoing large deformation along with an efficient and stable numerical update scheme. Numerical results showed convergence for varying mesh size and increasing regularization parameter $\beta$. For a smooth iterative solution scheme, it was necessary to develop a stabilization technique that allows for erosion of finite elements with damage exceeding a critical damage state without the need of remeshing. Concluding, we obtained an efficient numerical approach for damage processes to be described in the large deformation setting. In future investigations, important physical aspects like plasticity and hardening will be included.

\section*{Acknowledgment} 
The authors J. Riesselmann and D. Balzani greatly appreciate funding by the German Science Foundation (Deutsche Forschungsgemeinschaft, DFG) as part of the Priority Program German 1748 ``Reliable simulation techniques in solid mechanics. Development of non-standard discretization methods, mechanical and
mathematical analysis'', project ID BA2823/15-1.

\addcontentsline{toc}{chapter}{Bibliography}
\bibliographystyle{unsrt}
\bibliography{bib_short}

\end{document}